\documentclass[aps,pra,twocolumn,twoside,superscriptaddress]{revtex4-1}

\usepackage{amssymb}
\usepackage{cancel}
\usepackage{color,graphicx,bm}
\usepackage{amsmath}
\usepackage{amsbsy}
\usepackage{amsthm}
\usepackage[caption=false]{subfig}
\usepackage{bbm,float}
\usepackage{epsfig}
\usepackage{float}
\usepackage{graphicx}
\usepackage{dcolumn}
\usepackage{bbm}
\usepackage{color,epstopdf,dsfont}
\usepackage{amscd}
\usepackage{amsfonts}  
\usepackage[]{amsmath}
\usepackage{amssymb}    
\usepackage{mathrsfs}
\usepackage{verbatim}
\usepackage[]{cases}
\usepackage{amsmath}
\usepackage{wasysym}
\usepackage[latin9]{inputenc}
\usepackage[T1]{fontenc}
\usepackage{mathtools}
\usepackage{todonotes}
\usepackage[normalem]{ulem}
\usepackage[colorlinks=true,linkcolor=blue,urlcolor=blue,citecolor=blue,pdfusetitle]{hyperref}
\usepackage{cleveref}
\usepackage{libertine}

\usepackage{hyperref}
\usepackage{xr-hyper}
\newtheorem{theorem}{Theorem}

\newtheorem{corollary}[theorem]{Corollary}

\newcommand{\MP}[1]{\textcolor{black}{#1}}
\newcommand{\GGrev}[1]{\textcolor{black}{#1}}

\newcommand{\tr}{\operatorname{\bf{tr}}} % trace of a matrix
\newcommand{\ket}[1]{\vert #1 \rangle} % ket vector
\newcommand{\bra}[1]{\langle #1 \vert} % bra vector
\newcommand{\cM}{{\cal M}}

\renewcommand{\vec}[1]{\text{\boldmath$#1$}}

\newcommand{\ketbra}[2]{\ensuremath{| #1 \rangle\!\langle #2 |}}

\newcommand{\re}{\mathrm{Re}}

\newcommand{\be}{\begin{equation}} 							
\newcommand{\ee}{\end{equation}}
\newcommand{\bea}{\begin{eqnarray}}
\newcommand{\eea}{\end{eqnarray}}
\newcommand{\bematrix}{\left(\begin{matrix}}
\newcommand{\ematrix}{\end{matrix}\right)}
\def\tr{\mathrm{Tr}}

\definecolor{ao}{rgb}{0.0, 0.5, 0.0}

\def\En{E^{0}_n}
\def\Em{E^{\tau}_m}

\def\cM{\mathcal M}

\def\cM{\mathcal M}

\def\tr{\mathrm{Tr}}

\begin{document}
\title{Quantum work statistics with initial coherence}

\author{Mar\'ia Garc\'ia D\'iaz}
\affiliation{F\'{\i}sica Te\`{o}rica: Informaci\'{o} i Fen\`{o}mens Qu\`{a}ntics, %
	Departament de F\'{\i}sica, Universitat Aut\`{o}noma de Barcelona, ES-08193 Bellaterra (Barcelona), Spain}

\author{Giacomo Guarnieri}
\affiliation{School of Physics,  Trinity College Dublin,  College Green,  Dublin 2,  Ireland}

\author{Mauro Paternostro}
\affiliation{Centre  for  Theoretical  Atomic,  Molecular,  and  Optical  Physics,School of Mathematics and Physics, Queen's University, Belfast BT7 1NN, United Kingdom}

\begin{abstract}

The Two Point Measurement scheme for computing the thermodynamic work performed on a system requires it to be initially in equilibrium. The Margenau-Hill scheme, among others, extends the previous approach to allow for a non-equilibrium initial state. We establish a quantitative comparison between both schemes in terms of the amount of coherence present in the initial state of the system, as quantified by the $l_1$-coherence measure. We show that the difference between the two first moments of work, the variances of work and the average entropy production obtained in both schemes can be cast in terms of such initial coherence. Moreover, we prove that the average entropy production can take negative values in the Margenau-Hill framework.  
\end{abstract}

\maketitle

\section{Introduction}

In the quest for the understanding of the interplay between thermal and quantum fluctuations that determine the energy-exchange processes occurring at the nano- and micro-scale, the identification of the role played by quantum coherences is paramount~\cite{BookQT,deffner2019qtd}. The foundational nature of such understanding has been the driving force for much research effort, which has started shedding light onto the role that quantum coherence has in the quantum thermodynamic phenomenology, from work extraction to the emergence of irreversibility~\cite{Santos2019,FrancicaPRE2019,Riechers2020,Sone2020,Francica2020,Francica2020bis,Francica2019bis,Bernards2019,miller2020thermodynamic,Miller_2019,sc2019quantum}. Owing to the success that it has encountered in classical stochastic thermodynamics, the current approach to the determination of the statistics of such energetics in the quantum domain is based on the so-called two-point measurement (TPM) protocol~\cite{TalknerPRE2007,CampisiRMP2011,EspositoRMP2009}: the energy change of a system driven by a time-dependent protocol is measured both at the initial and final time of the dynamics. The application of the TPM protocol has led to the possibility to address the statistics of quantum energy fluctuations in a few interesting experiments~\cite{Batalhao2014,an2015experimental,peterson2019experimental,ronzani2018tunable,FlyWheel}. Unfortunately, such strategy has a considerable drawback in that, by performing a strong initial projective measurement, all quantum coherences in the energy eigenbasis are removed, {\it de facto} washing out the possibility of quantum interference to take place. 

This fundamental bottleneck has led to efforts aimed at formulating coherence-preserving protocols for the quantification of the statistics of energy fluctuations resulting from a quantum process~\cite{allahverdyan2014,Micadei2020,levy2019,Gherardini2020}. A particularly tantalising one entails the use of quasi-probability functions to account for such statistics~\cite{allahverdyan2014,Solinas2015,Solinas2016}. Drawing from the success that quasi-probability distributions have in signalling non-classical effects in the statistics of light fields, Ref.~\cite{allahverdyan2014,levy2019} have put forward the cases for the Margenau-Hill (MH) quasi-probability distribution~\cite{Terletsky1937,Margenau1961} for the energetics of a quantum process. The MH distribution, which is the real part of the well-known complex Kirkwood distribution~\cite{Kirkwood1933}, provides the probability distribution for any two non-commuting observables and can take negative values. In the context of stochastic thermodynamics, the distribution of energy fluctuations provided by the MH approach generalizes the TPM one by replacing the strong initial measurement requested by the latter with a weak measurement. 

Negative values of the statistics inferred following the ensuing protocol witness strong non-classicality of the overall process followed by the system~\cite{levy2019}, which are completely removed from the picture provided by TPM. In such a context, it is crucial to pinpoint the role that the quantum coherences either present in the initial state of the system or created throughout its dynamics have in the setting up of the MH phenomenology. This is precisely the point addressed in this paper, where we thoroughly investigate the differences between the statistics entailed by the TPM and MH approaches and relate them to the value taken by well-established quantifiers of quantum coherence~\cite{baumgratz2014} over the initial state of the system, as well as dynamical features of the process that the latter undergoes. We show that such coherence-depending differences have strong implications for the formulation of statements on the degree of irreversibility of a non-equilibrium process provided by the MH approach, and provide a re-formulation of the average entropy production that clearly highlights the contribution resulting from quantum coherences. %\MP{more details? highlight qudit-vs-qubit? I think it's not necessary at this stage, but won't be against it if you like to add more} \GGrev{I would actually prefer to keep it like this, without further details which would risk to hinder the message} 
This work thus makes the first, necessary steps towards the quantitative understanding of the implications of quantum coherence for the phenomenology of the statistics of energy fluctuations in the quantum domain. 

The remainder of this paper is organized as follows: In Sec.~\ref{statistics} we present a detailed description of the TPM and MH schemes. Sec.~\ref{coherence} is a brief introduction to coherence theory. The distance between the \GGrev{first} moments of work obtained in both schemes is related to initial coherence throughout Section \ref{distance_moments}. A similar investigation is carried out for the variances of work and the average entropy production, which can be found in Sec.~\ref{distance_variances} and \ref{study_entropy}, respectively. In Sec.~\ref{conclusions} we draw our conclusions, while we defer a series of technical details, including the demonstration of the main results of our work, to the accompanying Appendix.

\section{Background}
\subsection{Quantum work statistics}\label{statistics}
Consider an isolated quantum system initially prepared in an equilibrium state and subjected to an external force that changes a work parameter $\lambda_t$ in time according to a generic finite-time protocol. The latter includes, at the initial time $t=0$ and final time $t=\tau$, projective measurements of the energy of the system, which result in the values $E^{\lambda_0}_n\equiv E^0_n$ and $E^{\lambda_\tau}_m\equiv E^\tau_m$. Here, $n$ and $m$ label the respective energy levels of the initial and final Hamiltonian $H(\lambda_0)\equiv H_0$, $H(\lambda_\tau)\equiv H_\tau$ of the system. Thermal and quantum randomness render the measured energy difference $E^{\tau}_m-E^{0}_n$, which can be interpreted as the work done on the system through the protocol, a stochastic variable. One can recognize here the well-known Two Point Measurement (TPM) scheme for measuring work, whose values are distributed according to the following probability distribution:
\begin{equation}
p^{\text{TPM}}_\tau(w)=\sum_{m,n}P^{\text{TPM}}_\tau[\Em,\En] \delta[w-(\Em-\En)].
\end{equation}
Here $P^{\text{TPM}}_\tau[\Em,\En] $ is the joint probability to measure the energy values $\En$ and $\Em$, 
\begin{equation}
P^{\text{TPM}}_{\tau}[\Em,\En]=\tr \left[\Pi_{\Em} U_\tau \Pi_{\En} {\cal G}_0 \Pi_{\En} U_\tau^\dagger \Pi_{\Em} \right],
\end{equation}
where ${\cal G}(\lambda_t)\equiv{\cal G}_t={e^{-\beta H_t}}/Z_t$ is a Gibbs state---at the inverse temperature $\beta$---of the instantaneous Hamiltonian $H(\lambda_t)\equiv H_t=\sum_i E_i^{t}\, \Pi^t_i$, $Z_t=\tr\left[e^{-\beta H_t}\right]$ is the associated partition function, $\Pi_i^{t}=\ketbra{E_i^{t}}{E_i^{t}}$ is the projector onto the eigenstate $\ket{E^t_i}$ of $H_t$ with energy $E^t_i$, and $U(\tau)\equiv U_t$ is the unitary propagator of the evolution. 

Suppose now that our initial system was instead in a non-equilibrium state of the form $\rho^{\text{ne}}={\cal G}_0+\sum_{i\neq j}\rho^{\text{ne}}_{ij}$. It can be noticed that $p^{\text{TPM}}_\tau(w)$ would remain invariant in this case, since the action of the first projective measurement, performed through $\Pi^0_n$, destroys any coherence that could be present in the initial state. The following question can then be posed: what alternative protocols could be devised such that the initial-state coherence would have an effect on the measured thermodynamic work? 

Several strategies beyond the TPM scheme have been pointed out in this line~\cite{levy2019,allahverdyan2014,Micadei2020,Francica2020,miller2017,Gherardini2020,Sone2020,Riechers2020}. Here we will consider the MH scheme for measuring work, which replaces the first projective measurement of the TPM scheme with a weak measurement \cite{levy2019}, and thus allows for initial coherence to survive along the protocol. For an initial state $\rho^{\text{ne}}$, the values of work are now distributed according to
\begin{equation}
p^{\text{MH}}_\tau(w)=\sum_{m,n}P^{\text{MH}}_\tau[\Em,\En] \delta[w-(\Em-\En)],
\end{equation}
where 
\begin{equation}
P^{\text{MH}}_{\tau}[\Em,\En]=\re\left(\tr [U_t^\dagger \Pi_{\Em} U_t \Pi_{\En} \rho^{\text{ne}} ]\right)
\end{equation}
 is the MH quasiprobability distribution, which can take negative values in the range $P^{\text{MH}}_\tau[\Em,\En]\in [-{1}/{8},1]$ when the state that we consider deviates from equilibrium~\cite{allahverdyan2014}, and goes back to the distribution associated with a TPM approach for initial equilibrium states.  
 
\subsection{Coherence theory}\label{coherence}
The coherence of a state can be cast within the framework set by the well-established resource theory of coherence \cite{aberg2006,BraunGeorgeot,baumgratz2014,streltsov2017,winter2016}. As in every quantum resource theory \cite{brandao2015a}, free states and operations must be first identified: here the set ${\cal I}$ of free states---denoted as \emph{incoherent states}---includes all the states $\delta\in\mathcal{S(H)}$ \GGrev{(with $\mathcal{S(H)}$ denoting the set of unit-trace and semi-positive definite linear operators on $\mathcal{H}$)} that are diagonal in 
some fixed basis $\{\ket{i}\}_{i=0}^{d-1}$ of $\mathcal{H}$, whereas free operations are those that map the set of free states to itself and thus cannot generate coherence. The largest class of free operations are the 
\emph{maximally incoherent operations} (MIOs)~\cite{aberg2006}, 
consisting of all completely positive and trace-preserving (CPTP) maps $\cM$ such that $\cM({\cal I})\subset {\cal I}$. A subset of MIOs 
are \emph{incoherent operations} (IOs)~\cite{baumgratz2014}, comprising all CPTP maps $\cM$ that admit a Kraus representation 
with operators $K_\alpha$ such that $K_\alpha {\cal I} K_\alpha ^\dagger \subset {\cal I}$ 
for all $\alpha$. Only after singling out the states and operations that can be performed at no cost, can one investigate how resource states---states with coherence---are to be quantified, manipulated, and interconverted among each other. \emph{Coherence measures}~\cite{baumgratz2014} are indispensable at this stage: quantifying the amount 
of coherence present in a state $\rho\in\mathcal{S(H)}$, a coherence measure is a functional \GGrev{
$C:\mathcal{S(H)}\to\mathbb{R}_{\geq0}$ that fulfills the following conditions: (i) faithfulness, meaning that $C(\delta)=0$ for all $\delta\in{\cal I}$, and
(ii) monotonicity, $C(\rho)\geq C(\cM(\rho))$, for all free operations $\cM$. }% (iii) strong-monotonicity, $\sum_i q_i C(\rho_i) \leq C(\rho)$, where $\rho_i = K_i\rho K^{\dagger}_i/q_i$ denote the post-measurement states, $q_i = \mathrm{Tr}\left[K_i\rho K^{\dagger}_i\right]$ are the corresponding probability and $K_i$ are an incoherent Kraus operator; (iv) convexity $\sum_i p_i \mathcal{C}(\rho_i) \geq \mathcal{C}\left(\sum_i p_i \rho_i \right)$.}

\begin{figure}
\centering
\includegraphics[width=0.5\linewidth]{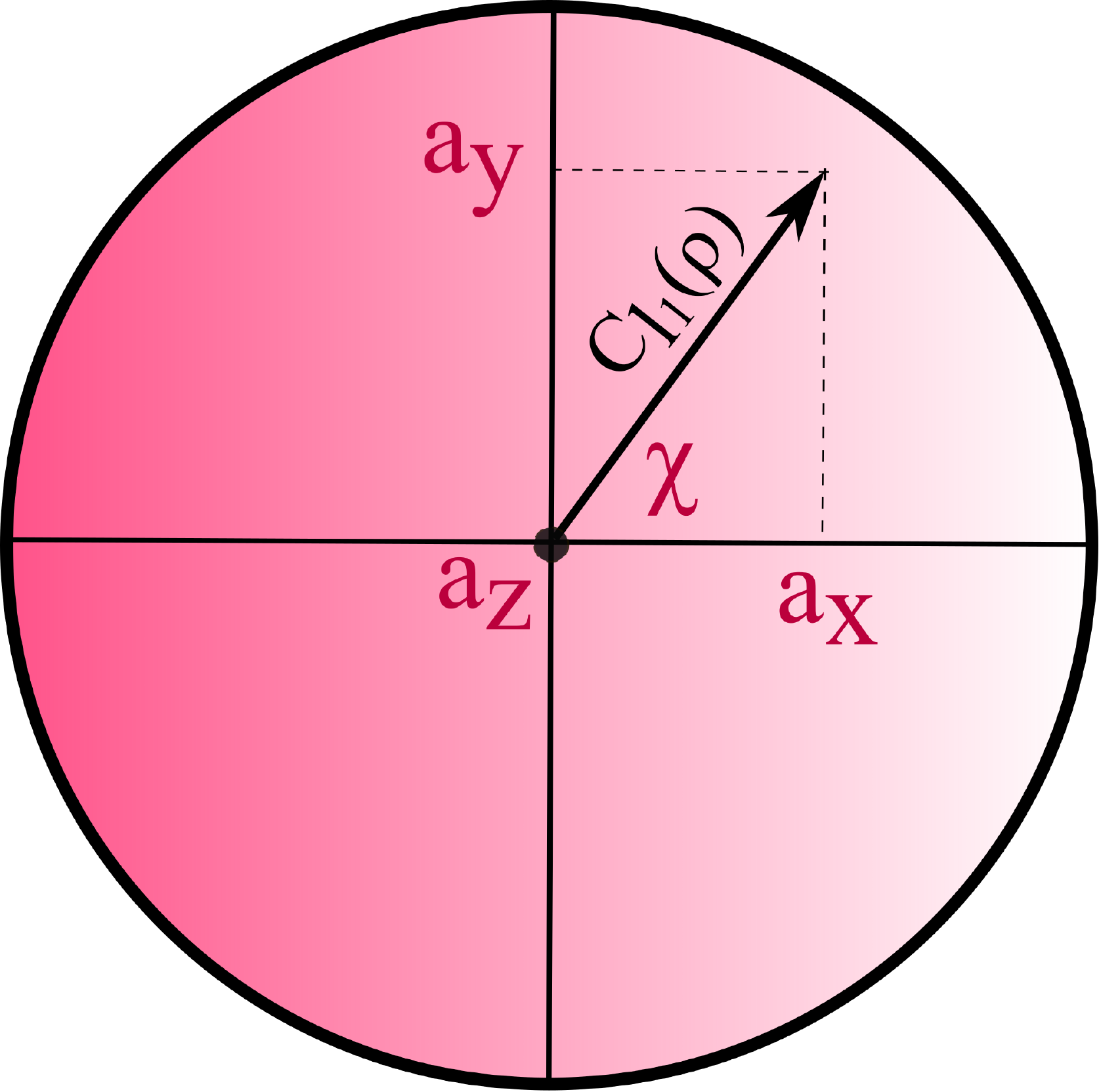}
\caption{Equatorial plane of the Bloch sphere at $z=0$. The $l_1$-coherence of a state quantifies its distance from the $z$-axis.}
\label{fig:bloch}
\end{figure}

In particular, throughout this work we will make use of the $l_1$-coherence measure \cite{baumgratz2014} defined as 
  \begin{equation}
  \label{Cmeas}
C_{l_1}(\rho)=\sum_{i\neq j} |\rho_{ij}|,
  \end{equation}
   which is a valid coherence quantifier under IOs, but not MIOs \cite{bu2017b}. Notably, when used on qubit states parametrized as $\rho=\frac{1}{2}(\mathds{1}+\vec{a}\cdot{\bm\sigma})$ %$\rho=\frac{1}{2}\begin{pmatrix}
%1-a_z & a_x-i a_y \\
%a_x+i a_y & 1+a_z 
%\end{pmatrix}$,
with $\vec{a} \in \mathbb{R}^3$ the Bloch vector associated with $\rho$ and ${\bm\sigma}$ the vector of Pauli matrices, Eq.~\eqref{Cmeas} quantifies the length of the projection ${\bm a}_\perp$ of ${\bm a}$ onto the equatorial plane of the Bloch sphere. % to the $z$-axis of the Bloch sphere: $C_{l_1}(\rho)=\sqrt{a_x^2+a_y^2}$. 
Thus, for qubit states with \GGrev{$a_z=0$}, we have $a_x=C_{l_1}(\rho)\cos(\chi)$ \GGrev{and $a_y=C_{l_1}(\rho)\sin(\chi)$} , where $\chi$ is the angle between ${\bm a}_\perp$ and the $x$-axis of the Bloch sphere (cf. Fig. \ref{fig:bloch}).

 \section{Main results}
 As previously stated, the purpose of this work is to provide a quantitative connection between the TPM scheme and the MH one in terms of quantum coherence. As all the information about a distribution is encoded in its moments, our approach to the assessment of the link between both schemes will rely on quantifying the distance between their corresponding moments. 
 
\subsection{Distance between the averages of work}\label{distance_moments}

The generating function of $p^{\cal O}_\tau(w)$ is defined as the Fourier transform $G_{{\cal O}}(\eta,\tau)=\int dw p^{\cal O}_\tau(w) e^{i\eta w}$ with ${\cal O}=\text{TPM},\text{MH}$. \GGrev{Moments of work are obtained through differentiation with respect to $\eta$, $\langle w^m_{\GGrev{\tau}} \rangle_{\cal O} =(-i)^k\frac{d^k}{d\eta^k}G_{{\cal O}}(\eta,\tau)\big|_{\eta=0}$~\cite{esposito2009}.}
In the TPM scheme, the latter can be written as
\begin{equation}
\langle w^m_{\GGrev{\tau}} \rangle_{\text{TPM}}=%\hspace{5.7cm} \nonumber\\
\tr \left[\Delta (\rho_{\GGrev{0}})(U_\tau^\dagger H_\tau U_\tau{-}H_0)^m\right],
\end{equation}
where $\rho_0$ is the initial state of the working medium and \GGrev{$\Delta(\rho_0)=\sum_n  \Pi^0_{n} \rho_0 \Pi^0_n $ is the fully dephasing map that suppresses coherences in the energy eigenbasis of the initial Hamiltonian (see notation in Sec.~\ref{statistics})}.
However, the corresponding quantity within the MH approach has the more involved form
 \begin{equation}
\langle w^m_{\GGrev{\tau}} \rangle_{\text{MH}}= %\hspace{5.8cm} \nonumber \\
 \frac{1}{2}\sum_{l=0}^m \binom{m}{l} \tr \left[ \left\{ H_\tau^l, (-H_0)^{m-l} \right \}  \rho_{\GGrev{0}}       \right],
 \end{equation} 
which reduces to $\langle w^m_\tau \rangle_{\text{MH}} = \tr\left[\rho_{\GGrev{0}} (U_\tau^\dagger H_\tau U_\tau-H_0)^m \right]$
only for $m=1,2$. It then becomes evident that the two first moments of work agree for both distributions whenever $[U_\tau^\dagger H_\tau U_\tau , H_0]=0$ or $[\rho_0, H_0]=0$~\cite{miller2017}.

If we then consider cyclic processes such that $H_0=H_\tau\equiv H=\sum_k h_k \ketbra{k}{k}$ %\GGrev{(I have removed everywhere the sentence `with a diagonal Hamiltonian` as it is always possible to diagonalize an Hamiltonian, being it an self-adjoint operator)}, 
we are led to our first result 
  \begin{theorem} \label{bound_qudit}
  For a $d$-dimensional system 
  %with a diagonal Hamiltonian $H=\sum_k h_k \ketbra{k}{k}$ fixing the basis w.r.t. which coherence is measured, and 
  undergoing a cyclic process described by a unitary evolution $U_\tau$, we have %the absolute difference between the first moments of work obtained via the MH scheme and the TPM scheme is upper-bounded as follows:
  \begin{equation}\label{qudit_maxU}
  |\langle w \rangle_{\textnormal{MH}}-\langle w \rangle_{\textnormal{TPM}}|\leq \frac{\tr|H|}{2}C_{l_1}(\rho_{\GGrev{0}}).
  \end{equation} 
The upper bound is tight for qubits, which are such that 
%  \begin{eqnarray}\label{qubit_dif}
 % |\langle w \rangle_{\textnormal{MH}}-\langle w \rangle_{\textnormal{TPM}}|=\hspace{4.3cm} \nonumber \\
 %\tr|H||\cos(\tau) \sin(\tau)\cos(\chi+\varphi_2-\varphi_1)| C_{l_1}(\rho)
  %\end{eqnarray}
  %and therefore
  \begin{equation} \label{max_qubit}
  \max_{U_\tau}|\langle w_{\GGrev{\tau}} \rangle_{\textnormal{MH}}-\langle w_{\GGrev{\tau}} \rangle_{\textnormal{TPM}}|=\frac{\tr|H|}{2}C_{l_1}(\rho_0),
  \end{equation}
where the maximum is sought over all unitary operations $U_\tau$. %qubit unitaries are generally parametrized as $U(\tau)=e^{i\frac{\varphi}{2}}\begin{pmatrix}
%  e^{i \varphi_1} \cos \tau & e^{i \varphi _2} \sin \tau \\
 % -e^{-i \varphi_2} \sin \tau & e^{-i \varphi_1} \cos \tau
 % \end{pmatrix}$ 
% and $|H|_{kk}:=|h_{kk}|$.
  \end{theorem}
%Theorem \ref{bound_qudit} reveals that bound (\ref{qudit_maxU}) is only tight for qubits. 
A proof is given in Appendix~\ref{appA}.
\GGrev{It is worth pointing out that the bound depends on initial time quantities, such as $C_{l_1}(\rho_0)$ as well as the Hamiltonian spectrum $\tr\left[|H|\right]$, thus clearly highlighting the role of initial coherences and the impact of the first initial projective measurement on them brought by the TPM scheme.}

The tightness of the bound in Eq.~\eqref{qudit_maxU} is quickly lost as the dimension of the information carrier grows. % this is no longer true, as 
For instance, Fig.~\ref{fig:maxUwork_vs_coh_3d} addresses the case of a system with $d=3$ showing the values taken by the exact (maximum) difference between the average work corresponding to the two strategies assessed here (red dots)---computed by means of random sampling---versus the degree of initial coherence in the state of the system. Such quantity is compared to the bound in Eq.~(\ref{qudit_maxU}) (blue crosses) to show a widening gap as $C_{l_{1}}(\rho_{\GGrev{0}})$ grows. However, a linear-like dependence with respect to the \GGrev{amount} of coherence can still be appreciated for the actual maximum distance between average works.
 
Let us get back to a qubit and consider the case of a sudden Hamiltonian quench, for which $U_\tau\rightarrow \mathds{1}$ \GGrev{in the limit $\tau\to 0$}. Under such conditions, we have $\langle w_{\GGrev{\tau\to 0}}\rangle_{\textnormal{MH}}=\langle w_{\GGrev{\tau\to 0}}\rangle_{\textnormal{TPM}}$,  irrespective of the initial coherence. This reflects the fact that both first moments of work will vanish individually under a sudden quench when considering cyclic processes, that is $\langle w_{\GGrev{\tau\to 0}}\rangle_{\text{TPM,MH}} \xrightarrow{U\rightarrow \mathds{1}} 0$. % and $\langle w\rangle_{\text{MH}} \xrightarrow{U\rightarrow \mathds{1}} 0$.\\

\subsection{Distance between the variances of work}\label{distance_variances}

\GGrev{The above analysis at the level of the averages of the work distributions is clearly insufficient to satisfactorily characterize the statistical implications of the first projective energy measurement which distinguishes between the TPM and MH schemes. It is in fact well known that measurements induce quantum fluctuations, which become extremely relevant whenever micro- and nano-scale systems are considered. Their connection with thermodynamics has recently drawn much attention and their role as a resource has been clarified~\cite{Ding2017,Auffeves2017,Buffoni_2019}.
} 
\GGrev{Driven by this, we now investigate the relationship between the variances of the work distribution in the TPM and MH schemes.
% in order to determine whether performing the initial projective energy measurement in the former protocol leads to an higher or smaller uncertainty. 
Somehow contrary to intuition, we will find that a definite general hierarchy between the two cannot be established, i.e. $( \Delta w_\tau)^2_{\textnormal{TPM}}$ is not greater or smaller than $( \Delta w_\tau)^2_{\textnormal{MH}}$ for all parameters. Instead, each particular experimental setup needs to be investigated on its own as either situation can occur.} 

\begin{figure}[t!]
\subfloat[\label{fig:maxUwork_vs_coh_3d}]{	\includegraphics[width=1\linewidth]{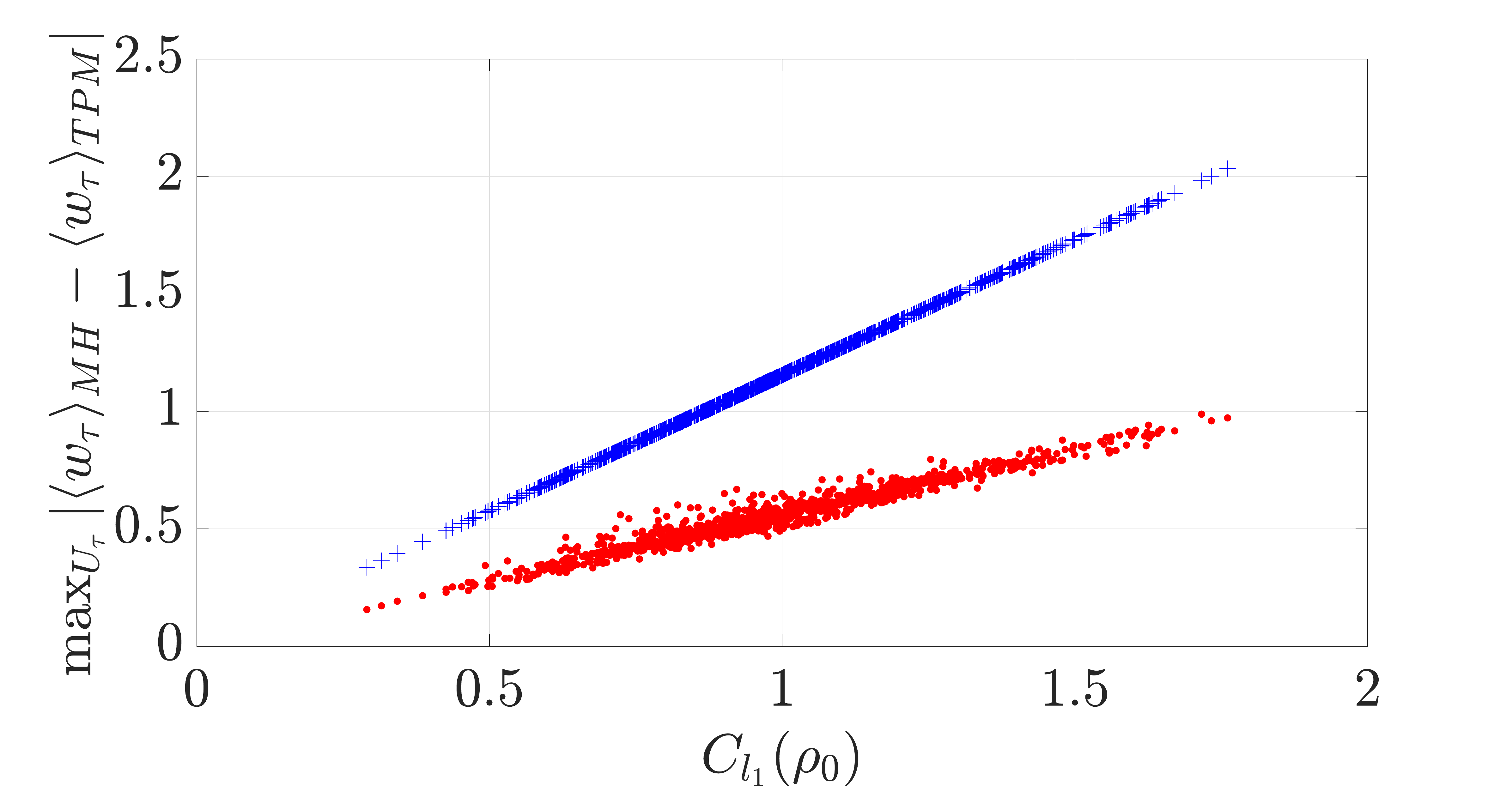}
}
	
\subfloat[\label{fig:maxUvarwork_vs_coh_3d}]{
	\includegraphics[width=1\linewidth]{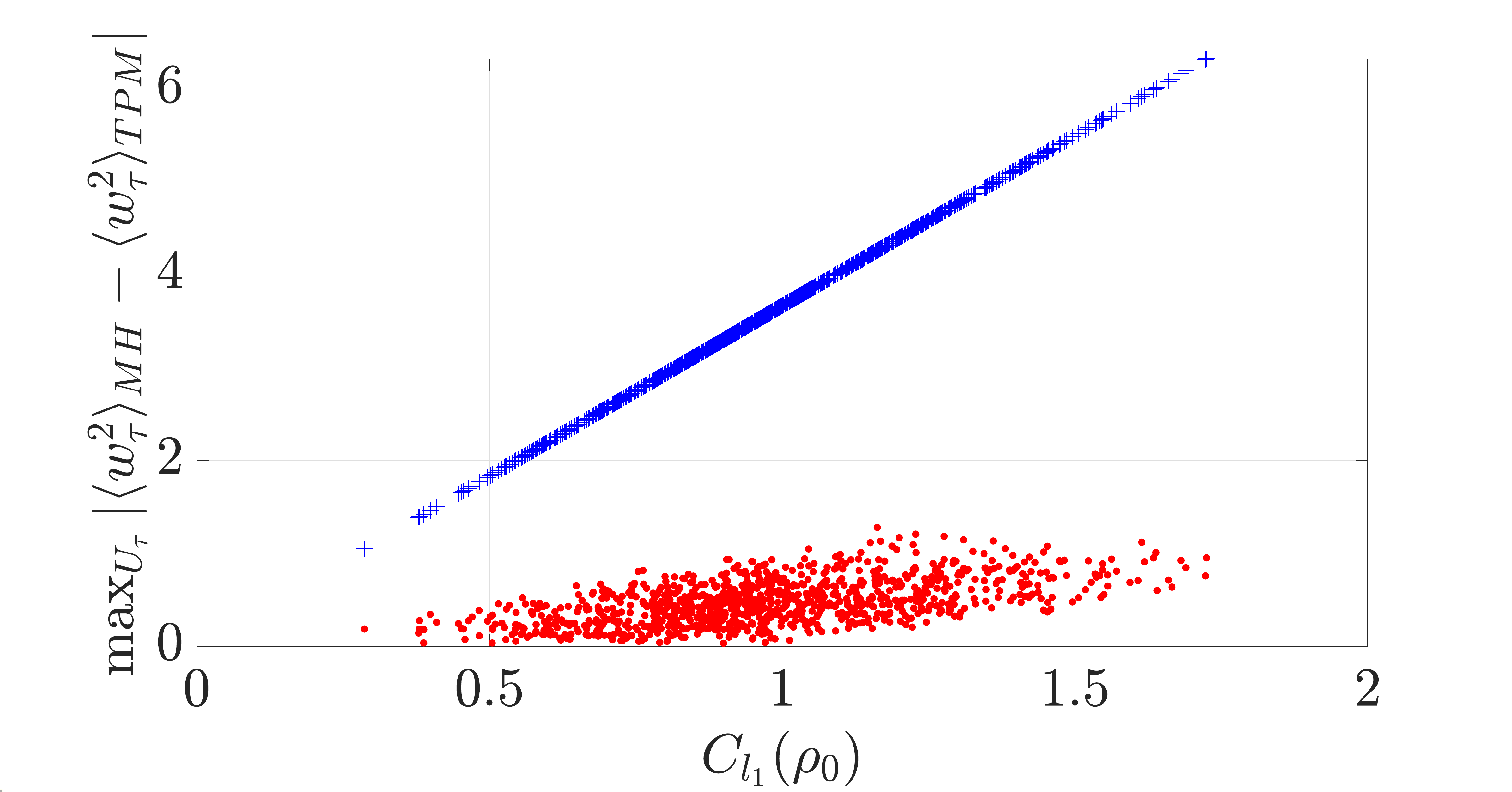}
}
\caption{\textbf{a)} Maximum absolute distance between the first moments of work obtained via the MH scheme and the TPM scheme (red dots) and the bound in Eq.~(\ref{qudit_maxU}) (blue crosses) versus initial coherence.
\textbf{b)} Maximum absolute distance between the second moments of work obtained via the MH scheme and the TPM scheme (red dots) and bound (\ref{bound_qudit_second}) (blue crosses) versus initial coherence. Both panels refer to a $d=3$ system governed by the Hamiltonian $H=(1/\sqrt{3})\text{Diag}[1,1,-2]$%\begin{pmatrix}
%	1 & 0 & 0 \\
%	0 & 1 & 0\\
%	0&0& -2
%	\end{pmatrix}$:
}
\end{figure}

Let us first of all focus on the second moment of the work distributions. In the same spirit of Theorem~\ref{bound_qudit}, we prove the following:
 \begin{theorem}\label{qudit_second}
 For a $d$-dimensional system 
 %with diagonal  Hamiltonian $H=\sum_k h_k \ketbra{k}{k}$ %fixing the basis w.r.t. which coherence is measured and 
 undergoing a cyclic process described by a unitary evolution $U(\tau)$, we have%, the absolute difference between the second moments of work obtained via the MH scheme and the TPM scheme is upper-bounded as follows:
 \begin{equation}\label{bound_qudit_second}
 |\langle w^2_{\GGrev{\tau}} \rangle_{\textnormal{MH}}{-}\langle  w^2_{\GGrev{\tau}} \rangle_{\textnormal{TPM}}|{\leq} % \hspace{3cm}  \nonumber\\
 \frac{C_{l_1}}{2}(\rho_0)\left(\tr H^2 {+} 2\max_k|h_k|\tr |H|\right).
 \end{equation} 
 When restricting the attention to qubits, we have
 \begin{equation}\label{second_qubit}
 |\langle w^2 \rangle_{\textnormal{MH}}-\langle w^2 \rangle_{\textnormal{TPM}}|=0.
 \end{equation}
 \end{theorem}
A detailed proof is reported in Appendix~\ref{appB}.
\GGrev{Once again, the bound Eq.~\eqref{bound_qudit_second} just depends on initial quantities such as the amount of coherences in the initial state $\rho_0$ and the energy spectrum of the initial Hamiltonian. However, at variance with the bound in Theorem~\ref{bound_qudit}, this is not tight even for qubits $d=2$, since the right-hand side of the inequality generally does not vanish and thus does not reduce to Eq.~\eqref{second_qubit}.}
\GGrev{In line with the analysis carried out for the} discrepancy between first moments, Fig.~\ref{fig:maxUvarwork_vs_coh_3d} illustrates the diverging gap between the bound in Eq.~(\ref{bound_qudit_second}) and the maximum difference between second moments for the case of a qutrit ($d=3$) system with a growing degree of quantum coherence in its initial state.  %bound in  is loose again in a given qutrit case. Moreover, we learn from Theorem \ref{qudit_second} that it is not even tight for qubits, since it does not vanish in general.  
  
\GGrev{Owing to Eq. (\ref{second_qubit}), the difference between variances can be simply calculated as  $( \Delta w_\tau)^2_{\textnormal{MH}}- ( \Delta w_\tau)^2_{\textnormal{TPM}}
=-\langle w_\tau\rangle^2_{\text{MH}}+\langle w_\tau\rangle^2_{\text{TPM}}$. These two considerations allow to show that the above difference does not have a definite sign in general. To show this, let us restict for simplicity to the case of a qubit undergoing an evolution described by
 \begin{equation}
 \label{Umat}
 U_\tau=\begin{pmatrix}
 \cos\tau & \sin\tau \\
 -\sin\tau & \cos\tau 
 \end{pmatrix}.
 \end{equation}
Then, it is straightforward to prove the following}
 \begin{corollary}\label{thm_variance}
 For a $d=2$ system undergoing a cyclic process described by a real unitary evolution $U_\tau$ %and a Hamiltonian $H=\sum_k h_k \ketbra{k}{k}$ fixing the basis w.r.t. which coherence is measured, 
 we have %the difference between the variances of work obtained via the MH scheme and the TPM scheme is given by
\begin{equation}
\begin{aligned}
( \Delta w_\tau)^2_{\textnormal{MH}}&- ( \Delta w_\tau)^2_{\textnormal{TPM}}=
\\& -f(\rho_0)[f(\rho_0)+2\sin(\tau)^2 a_z(h_0-h_1)],
\end{aligned}
\end{equation}
where $f(\rho_0)=(h_0-h_1)C_{l_1}(\rho_0)\sin(2\tau)\cos(\chi)/2$.
\end{corollary}

\begin{figure}

	\includegraphics[width=1\linewidth]{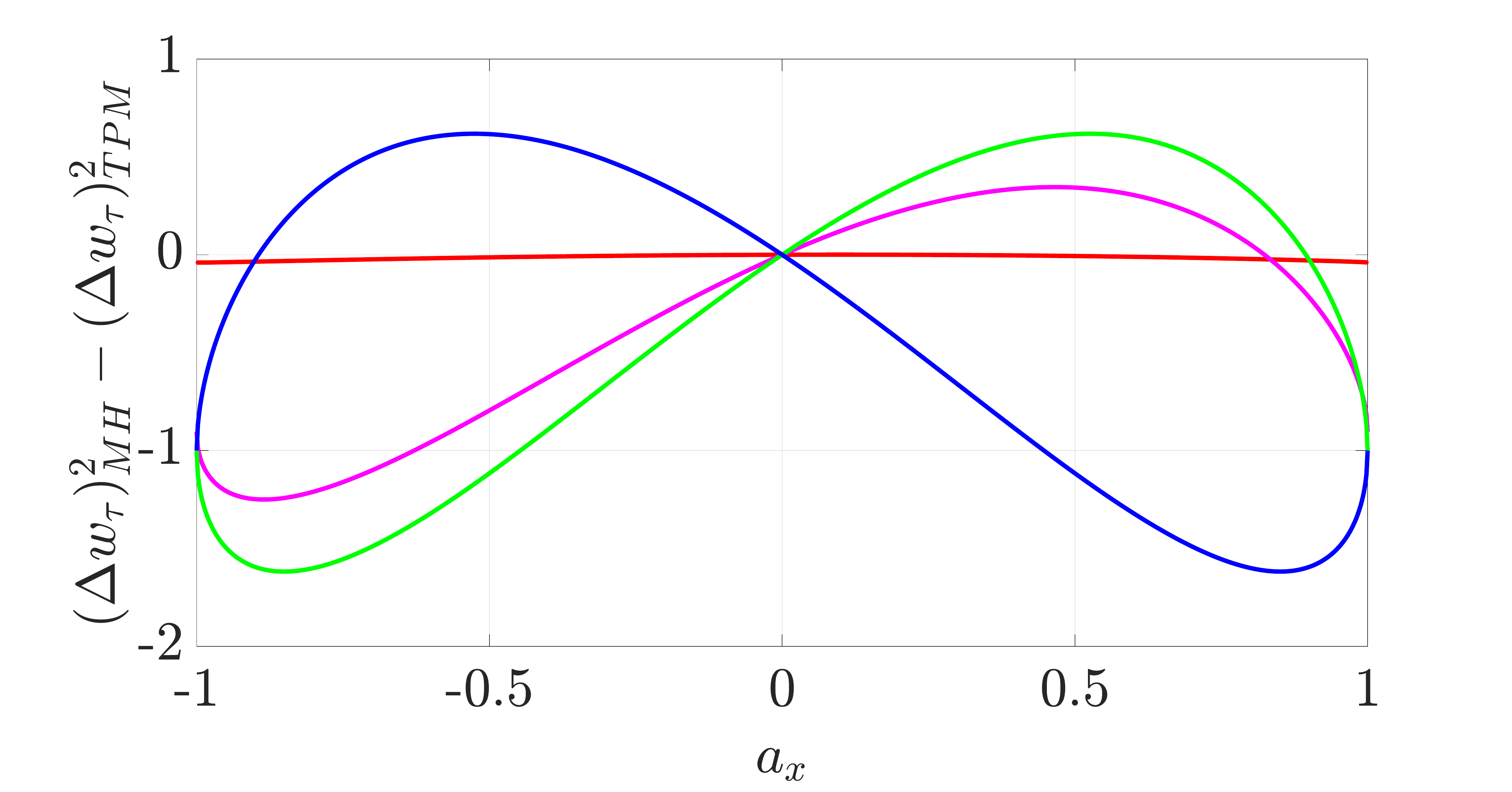}%

\caption{We plot the discrepancy $( \Delta w)^2_{\textnormal{MH}}- ( \Delta w)^2_{\textnormal{TPM}}$ between the variances of the TPM and MH distributions for pure states of $d=2$ systems with Bloch vector $(a_x,0,\sqrt{1-a_x^2})$ against $a_x$. We have taken $H_0=H_\tau=\sigma_z$ and the dynamics described by Eq.~\eqref{Umat}. We have taken $\tau=0.1$ (red), $\tau={\pi}/{5}$ (magenta), $\tau={\pi}/{4}$ (green) and $\tau={3\pi}/{4}$ (blue).
	\label{fig:difvars_realpure}}
\end{figure}

\begin{figure}[t!]
\subfloat[]{%
	\includegraphics[width=.5\linewidth]{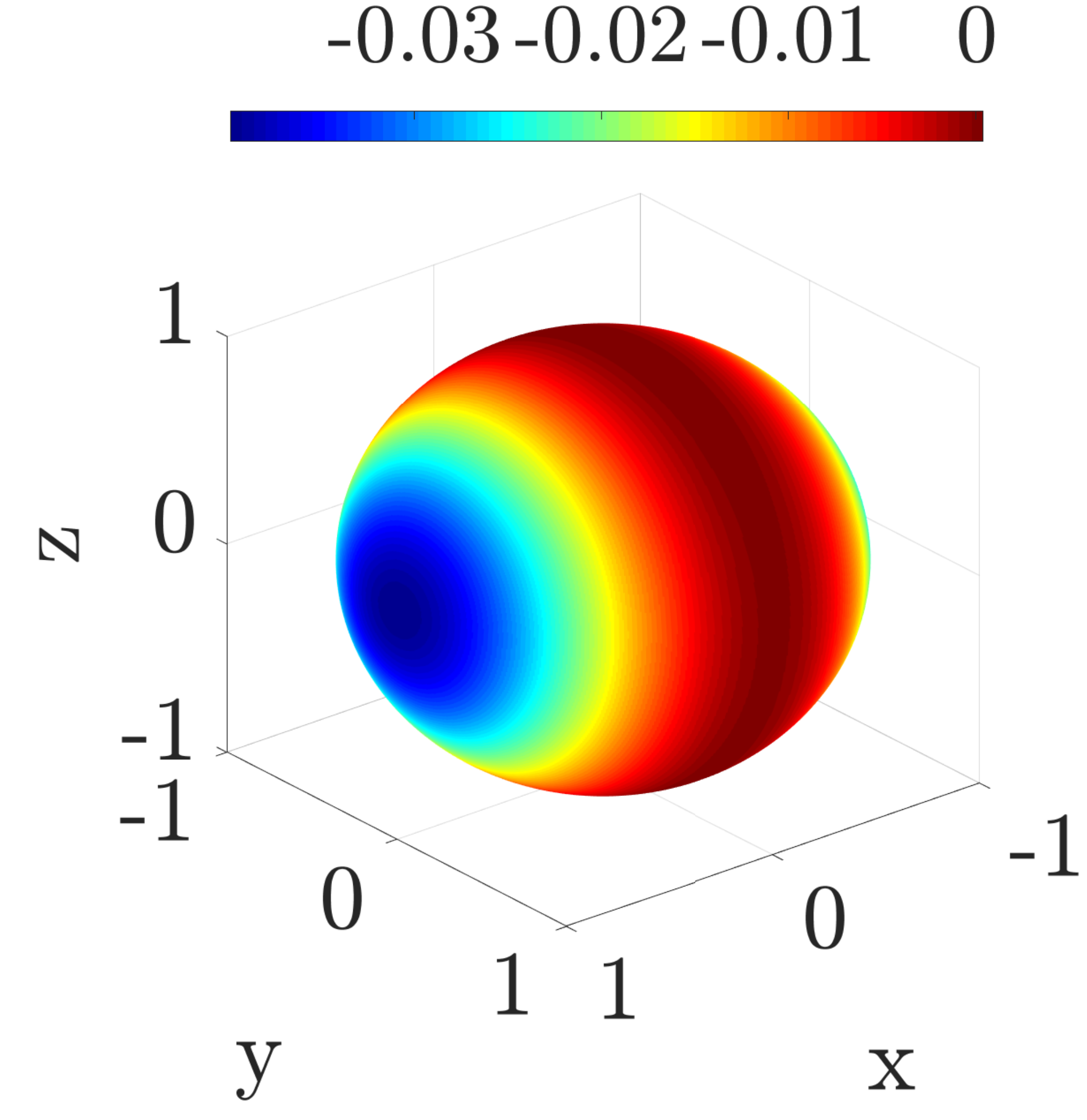}%
}
\subfloat[]{%
	\includegraphics[width=.5\linewidth]{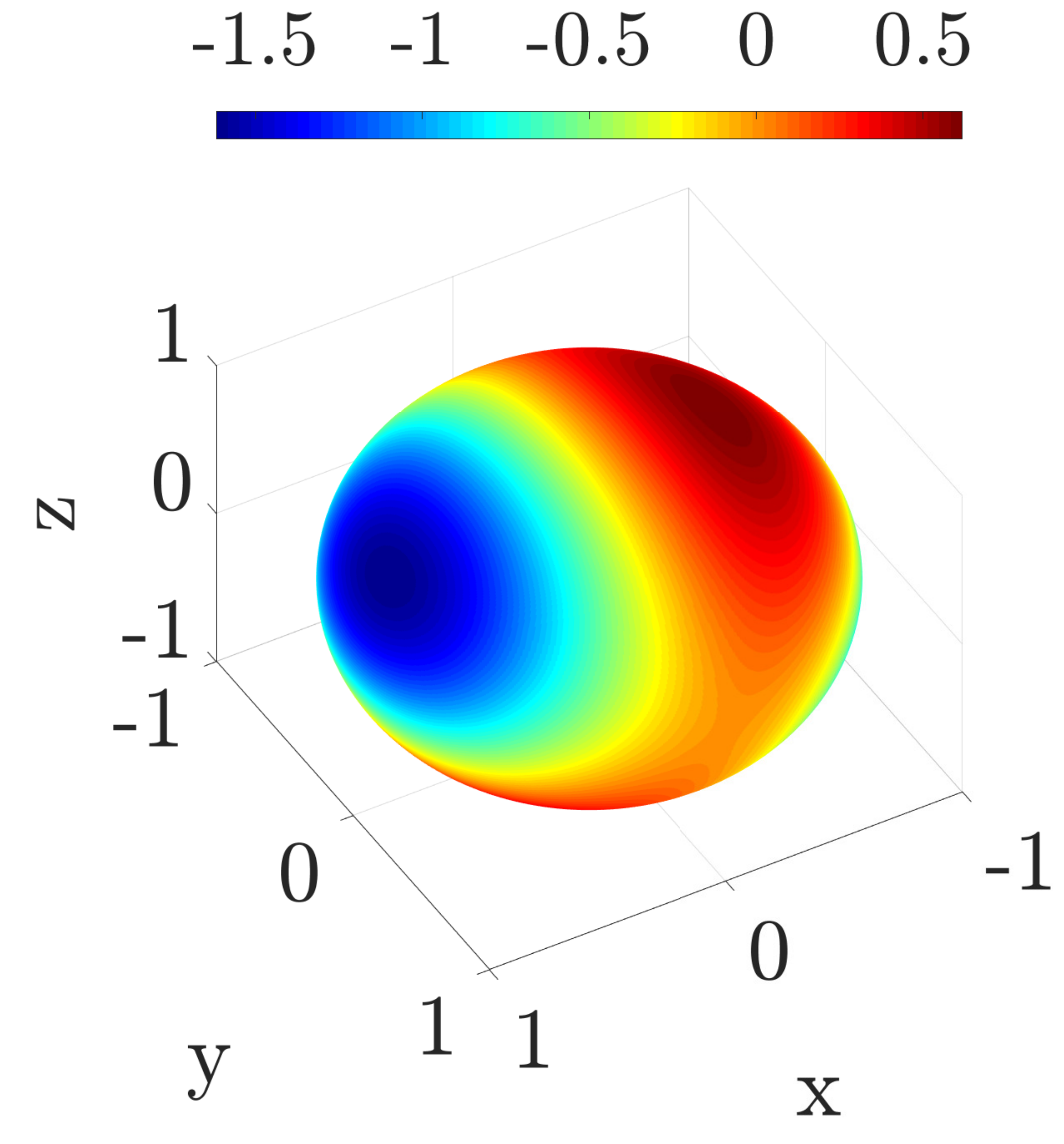}%
}\\
\subfloat[]{%
	\includegraphics[width=.5\linewidth]{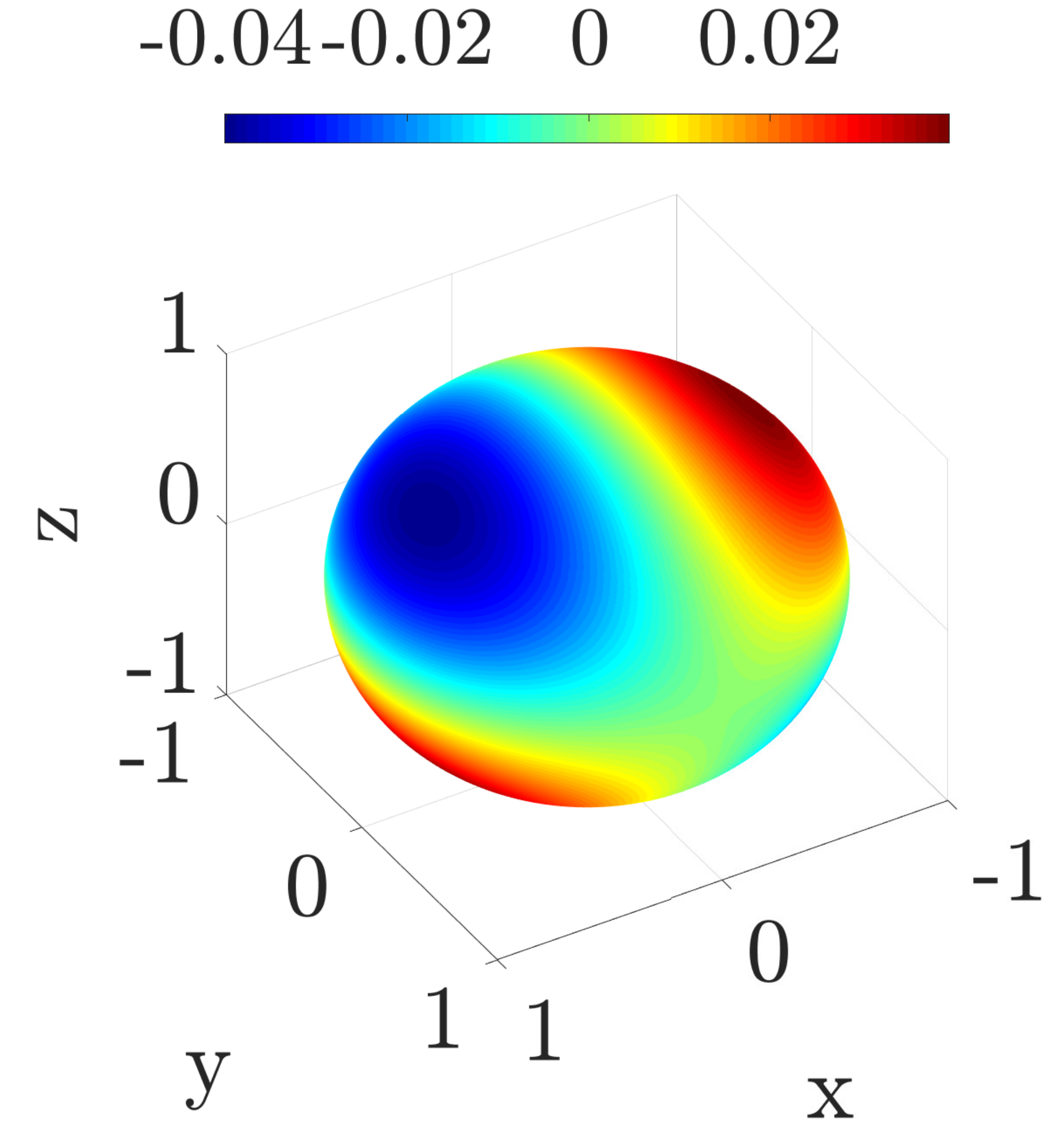}%
}
\subfloat[]{%
	\includegraphics[width=.5\linewidth]{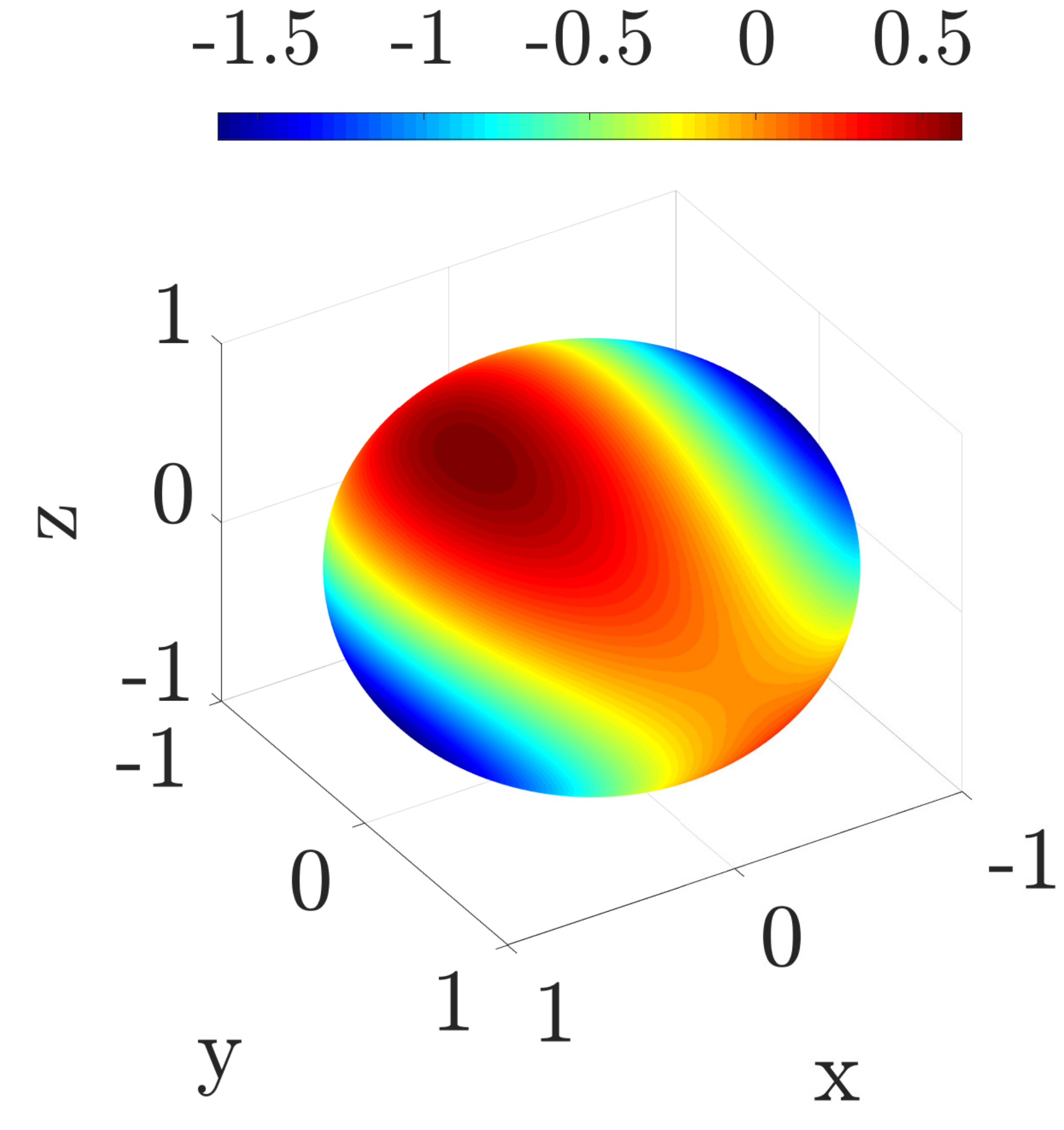}%
}
\caption{We correlate the value of the discrepancy $( \Delta w_\tau)^2_{\textnormal{MH}}- ( \Delta w_\tau)^2_{\textnormal{TPM}}$ between the variances of the TPM and MH distributions to the specific point on the Bloch sphere that represents a pure states of $d=2$ systems. We have used $H_0=H_\tau=\sigma_z$, the unitary propagator in Eq.~\eqref{Umat} and $\tau =0.1$ [panel {a)}], $\tau={\pi}/{4}$ [panel b)], $\tau={\pi}/{2}-0.01$ [panel c)] and $\tau={3\pi}/{4}$ [panel d)].
	\label{fig:difvars}}
\end{figure}

As we see in Figs. \ref{fig:difvars_realpure} and \ref{fig:difvars}, the difference between the variances can be either negative or non-negative, so it is not possible to determine which one is larger in general. Restricting to pure real qubits ($a_y=0 \rightarrow \cos(\chi)=\pm1$ and $a_z=\pm \sqrt{1-a_x^2}$), for easiness of the calculation, helps us discern which distribution is more uncertain depending on the values of $a_x$, as shown in Fig. \ref{fig:difvars_realpure} and summarized in Table \ref{fig:table} (further details about the corresponding analysis can be found in Appendix \ref{app_variances}). The results demonstrate that knowing the value of $C_{l_1}(\rho_0)$ does not suffice to ascertain which variance is larger: rather, it is the sign of $\cos(\chi)=\pm 1$ that eventually dictates their ordering. 

Considering the whole set of pure qubits ($a_y\neq 0\rightarrow \cos(\chi)\neq \pm1$ and $a_x^2+a_y^2+a_z^2=1$) would certainly require a much more involved analysis; however this exceeds the present purposes, which are just to point out that the contribution of $\cos(\chi)$, consistently with what was claimed for real qubits, can never be neglected when assessing the relative uncertainty between distributions (see Fig. \ref{fig:difvars}, where the difference between variances is shown to change significantly for different values of $\chi$).

\begin{table}[b]

\centering
\begin{tabular}{ccc}
 \hline
\hline
% $\varrho_{AB}$ & ${\cal S}(\varrho_{AB})$ & color\\
& $\text{sgn}(a_z)=\text{sgn}(\tan\tau)$ & \\
\hline\hline
    $a_x{\in}[-1,-2 a_z\tan\tau]$ & $a_x{\in}[-2 a_z \tan\tau,0]$ & $a_x\in[0,1]$\\ \hline \\[-1em]
$(\Delta w)^2_\text{MH}\le(\Delta w)^2_\text{TPM}$&$(\Delta w)^2_\text{MH}\ge(\Delta w)^2_\text{TPM}$&$(\Delta w)^2_\text{MH}\le(\Delta w)^2_\text{TPM}$\\ \hline
\end{tabular}\\
{a)}\\\vskip0.1cm
\begin{tabular}{ccc} 
\hline\hline
&{$\text{sgn}(a_z)\neq\text{sgn}(\tan\tau)$}&\\ 
\hline
\hline  
$a_x\in[-1,0]$ & $a_x\in[0,-2a_z\tan\tau]$& $a_x\in[-2a_z\tan\tau,1]$ \\  
\hline \\[-1em]
$(\Delta w)^2_{\text{MH}}\leq (\Delta w)^2_{\text{TPM}}$ & $(\Delta w)^2_{\text{MH}}\geq (\Delta w)^2_{\text{TPM}}$&$(\Delta w)^2_{\text{MH}}\leq(\Delta w)^2_{\text{TPM}}$ \\ 
\hline
\end{tabular}\\
{b)}
\caption{Relation between the variances of the MH and the TPM schemes, for a two-level system with a Bloch vector with $a_y=0$ and dynamics ruled by Eq.~\eqref{Umat}.
In table {a)} we have taken $\text{sgn}(a_z)=\text{sgn}(\tan\tau)$, while table {b)} is for $\text{sgn}(a_z)\neq \text{sgn}(\tan\tau)$.}
\label{fig:table}
\end{table}

%\begin{figure*}
%\centering
%\subfloat[]{\firsttab}%
%\\
%\subfloat[]{\secondtab}
%\caption{Relation between the variances of the MH and the TPM schemes, for real pure qubits and real qubit unitaries.
%\textbf{a)} $\text{sgn}(a_z)=\text{sgn}(\tan\tau)$.
%\textbf{b)} $\text{sgn}(a_z)\neq \text{sgn}(\tan\tau)$.
%}%
%\label{fig:table}%
%\end{figure*}

\subsection{Study of the entropy production}\label{study_entropy}

\begin{figure}[t!]
\subfloat[\label{fig:entropyvstimebeta02}]{%
	\includegraphics[width=1\linewidth]{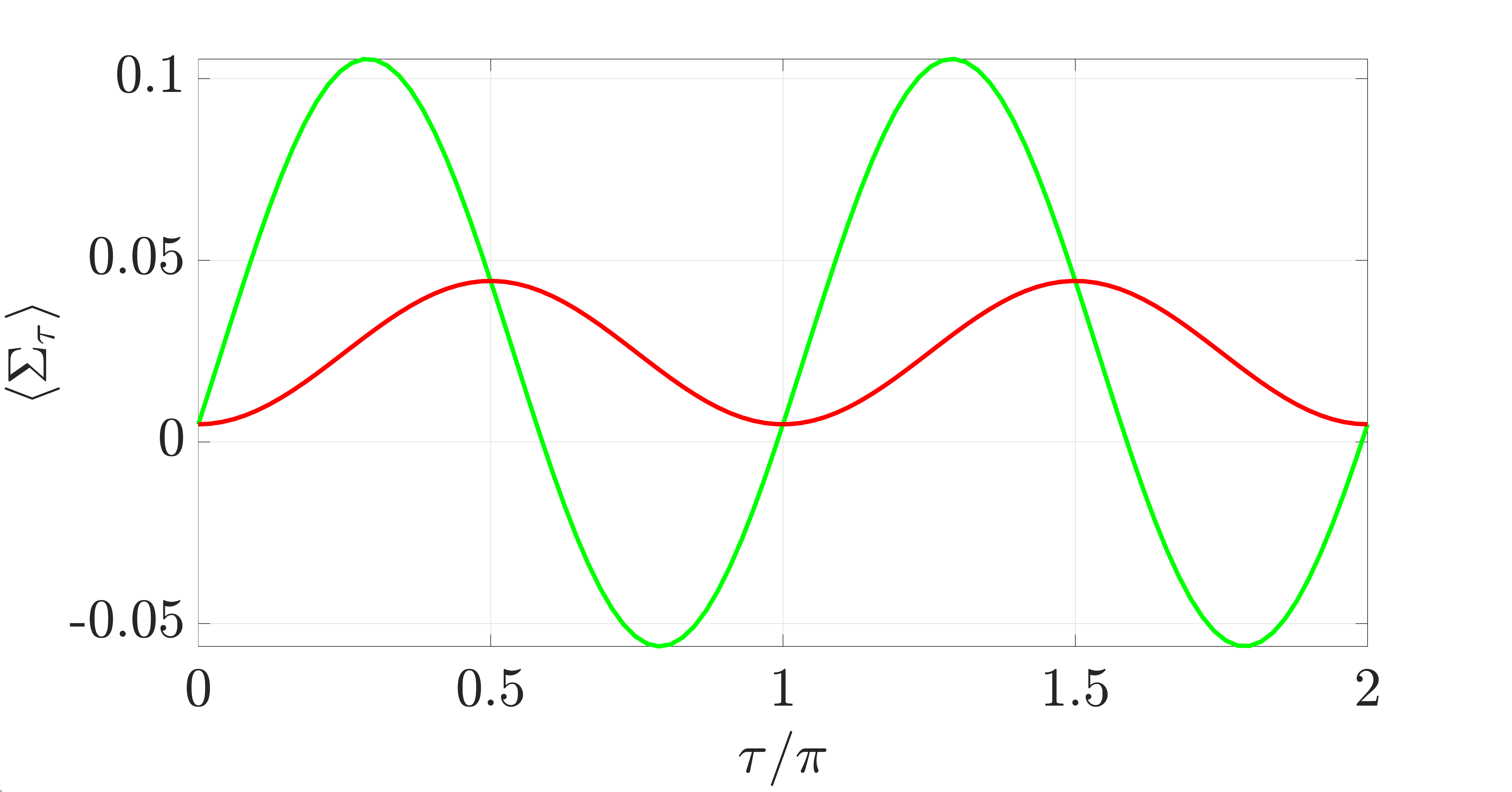}%
}
\\
\subfloat[\label{fig:entropyvsbetatau3pi4}]{%
	\includegraphics[width=1\linewidth]{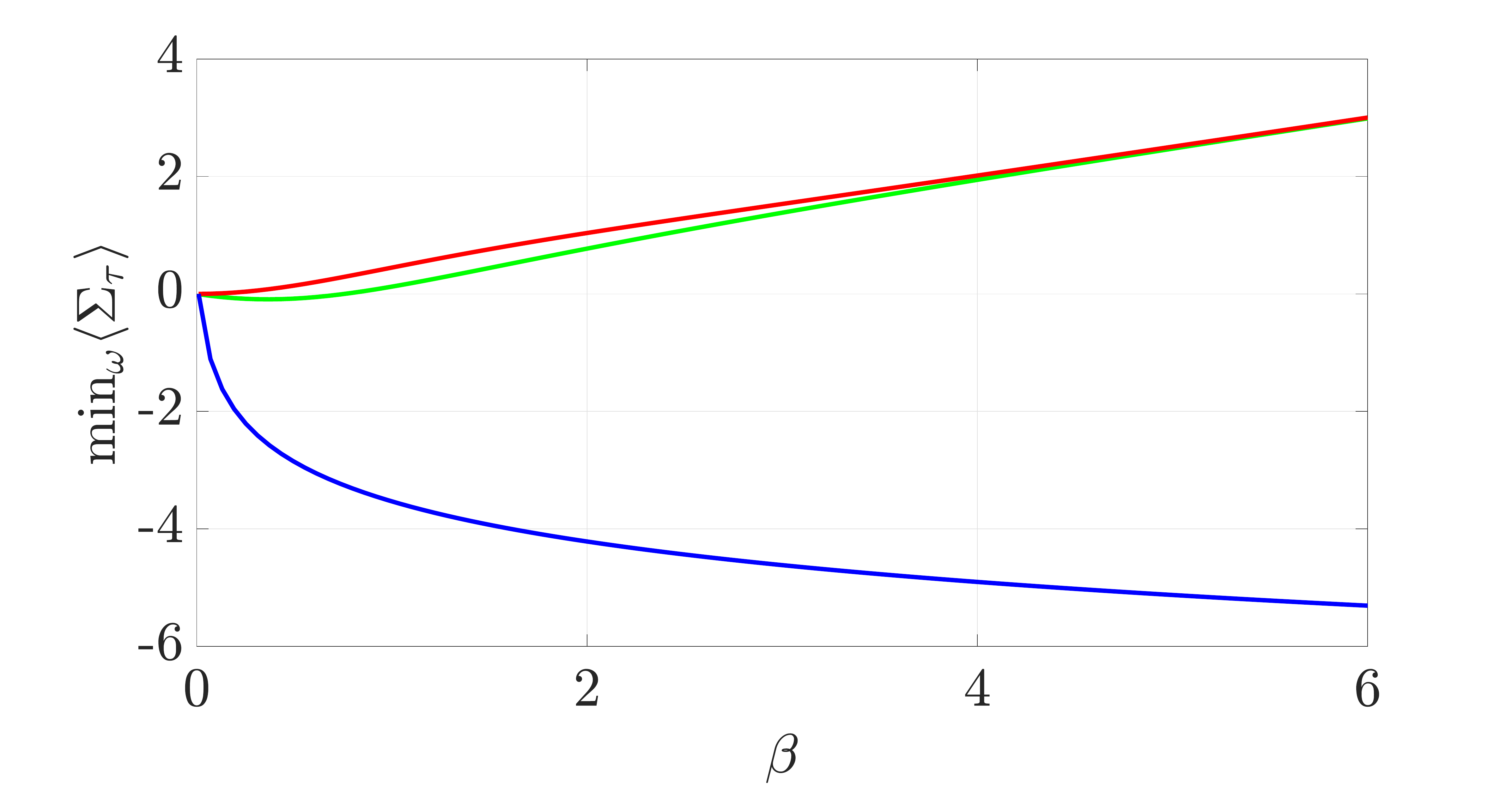}%
}
\caption{\textbf{a)} We plot the average entropy production $\langle \Sigma_\tau \rangle$ versus time for $\beta=0.2$, $\omega=0.8$. The behavior corresponding to the  MH (TPM) scheme  is shown by the green (red) curve.
	\textbf{b)} We show $\min_{\omega} \langle \Sigma_\tau \rangle$ for the MH and TPM schemes (green and red curves, respectively), and $-\log \xi$ (blue curve) versus $\beta$ for $\tau={3\pi}/{4}$. For both panels we have taken a qubit prepared in state $\rho=\begin{pmatrix}
	1-\alpha^2 & \omega \alpha \sqrt{1-\alpha^2}  \\
	\omega \alpha \sqrt{1-\alpha^2}  & \alpha^2 
	\end{pmatrix}$
	with $\alpha^2={e^{\beta }}/{\tr[e^{-\beta H_0}]}$, $\omega\in[0,1]$ and $H_0=\sigma_z$, undergoing an evolution given by a real unitary  ruled by Eq.~\eqref{Umat}
	%$U(\tau)=\begin{pmatrix}
	%\cos(\tau) & \sin(\tau) \\
	%-\sin(\tau) & \cos(\tau) 
	%\end{pmatrix}$
	with $H_\tau=\sigma_z/2$.
}
\label{fig:entropy_general}
\end{figure}

\GGrev{We conclude our analysis by exploiting the above results concerning the work statistics in order to investigate the consequences of initial coherence onto the second law of thermodynamics. In particular, whenever a system, initially prepared in a thermal state by contact with a bath at inverse temperature $\beta$, is then detached from it and unitarily brought out of equilibrium, then the work performed or extracted on the system is always on average greater or equal than the change in equilibrium free energy, i.e.
\begin{equation}\label{secondlaw}
    \langle \Sigma_\tau \rangle_{\text{TPM}} \equiv \langle w_\tau \rangle_\text{TPM} - \Delta F \geq 0
\end{equation}
with $\Delta F_\tau=\beta^{-1}\ln \frac{Z_0}{Z_\tau}$ and $Z_t = \mathrm{Tr}\left[e^{-\beta{H}_t}\right]$. The quantity $\langle \Sigma \rangle_{\text{TPM}}$ is also commonly known as dissipated work, or entropy production, and provides a measure of irreversibility of the work protocol.
Occasional violations to the second law can however take place due to the above mentioned work fluctuations.
Remarkably, the above inequality can be turned into an equality: this milestone result, known as Jarzynski equality ~\cite{jarzynski1997}, states that
\begin{equation}\label{eq:jarzynski}
\langle e^{-\beta (w_\tau-\Delta F_\tau)}\rangle_\text{TPM} = 1,
\end{equation}
from which Eq.~\eqref{secondlaw} is recovered by simple application of Jensen's inequality.
It is worth stressing that a key assumption behind the above results is to start with an initial state in thermal equilibrium $\rho_0 = \mathcal{G}_0 \equiv Z^{-1}_0 e^{-\beta\mathcal{H}_0}$. This state, which is clearly incoherent with respect to the initial Hamiltonian, implies that both the TPM and the MH schemes provide the same answer for the work distribution. We thus chose for convenience and clarity to use the subscript TPM in order to distinguish from the MH scenario when initial states with finite coherence in the energy eigenbasis are considered.}
For an arbitrary initial state $\rho_0$, in fact, the following fluctuation theorem has been shown to hold~\cite{allahverdyan2014}
\begin{equation}\label{eq:allahverdjan}
\langle e^{-\beta (w_\tau-\Delta F_\tau)}\rangle_{\text{MH}}= \re\left(\tr[\gamma_\tau {\cal G}_0^{-1}\rho_0]\right) \equiv \xi_\tau,
\end{equation}
where $\gamma_\tau \equiv  U_\tau^\dagger {\cal G}_\tau U_\tau$.
\GGrev{Eq.~\eqref{eq:jarzynski} is recovered for $\rho_0 = \mathcal{G}_0$. 
The consequences of the first projective measurement involved in the TPM scheme, whenever the system possesses initial coherence, can therefore be seen by comparing Eqs.~\eqref{eq:allahverdjan} and~\eqref{eq:jarzynski}.
In what follows we will in particular complement the analysis carried out in this respect in ~\cite{allahverdyan2014} by studying the average entropy production in the MH scheme. 
Thanks to the convexity of the function appearing in Eq.~\eqref{eq:allahverdjan}, one can still apply Jensen's inequality to obtain}
\begin{equation}
\langle \Sigma_\tau \rangle_{\text{MH}}=\beta(\langle w_\tau \rangle_{\text{MH}}-\Delta F_\tau)\geq -\ln \xi_\tau,
\end{equation}
which remarkably does not preclude a negative average entropy production (indeed, $\ln\xi_\tau$ can be arbitrarily large~\cite{allahverdyan2014}). 
This happens to hold for small enough $\beta$, as we can see in Figs.~\ref{fig:entropyvstimebeta02} and  Fig.~\ref{fig:entropyvsbetatau3pi4}, where we have plotted the minimum average entropy production as a function of $\beta$, for a suitable qubit evolution, both in the MH and the TPM schemes. From Fig. \ref{fig:entropyvsbetatau3pi4} we also note that both schemes seem to converge for $\beta \rightarrow \infty$, which is due to the fact that the coherence of $\rho$ gets smaller as $\beta$ increases. 

In order to get a deeper analytical insight of the regions where $\langle \Sigma_\tau \rangle_{\text{MH}}<0$, we go to the linear response regime~\cite{kubo1957statistical}.
Here we prove the following:
\begin{theorem}\label{mh_lr}
In the MH scheme, the average entropy production in the linear response regime amounts to
\begin{equation}\label{entropylr}
\begin{aligned}
\langle \Sigma_\tau \rangle_{\textnormal{MH}}^{\textnormal{LR}}&=\beta \langle w_\tau\rangle_{\textnormal{MH}} -\dfrac{\beta^2}{2}\re\left\{ \tr\left( \rho_0 [H_0,U_\tau^\dagger H_\tau U_\tau]\right)\right\} \\ 
&-\dfrac{\beta^2}{4}\tr[H_0^2-H_\tau^2].
\end{aligned}
\end{equation} 
For two-dimensional systems undergoing a process described by a real unitary evolution, $H_0=\sigma_z$ and $H_\tau=k\sigma_z$ ($k\in\mathbb{R}$), this yields
\begin{equation}\label{entropylrqubit}
\langle \Sigma_\tau \rangle_{\textnormal{MH}}^{\textnormal{LR}}-\langle \Sigma_\tau \rangle_{\textnormal{TPM}}^{\textnormal{LR}}=\beta k \sin(2\tau)\cos(\chi)C_{l_1}(\rho_0),
\end{equation}
where $\langle \Sigma_\tau \rangle_{\textnormal{TPM}}^{\textnormal{LR}}={\beta^2} ( \Delta w_\tau)^2_{\textnormal{TPM}}/2$.
\end{theorem} 
From Eq.~(\ref{entropylrqubit}) we see again that
both approaches are only equivalent when the initial state is in equilibrium (meaning that $C_{l_1}(\rho_0)=0$). Moreover, we notice that considering a cyclic process ($k=1$) would allow us to recover the relation between the first moments of work obtained in Theorem~\ref{bound_qudit}.
Finally, we observe that, under a sudden quench, both approaches agree irrespective of the initial state. This must be the case, since it has to be ensured that for $k=1$ (cyclic processes) both first moments of work vanish under a sudden quench, as argued in Section \ref{distance_moments}
\begin{equation}
\langle \Sigma_\tau \rangle=\beta\langle w_\tau\rangle =\beta \tr\rho(U_\tau^\dagger H U_\tau-H)\xrightarrow{U\rightarrow \mathds{1}} 0. \\
\end{equation}

We are now equipped to prove the achievability of $\langle \Sigma_\tau \rangle_{\text{MH}}<0$

\begin{corollary}\label{violation}
In the MH scheme, the average entropy production can take negative values, in contrast to what happens in the TPM scheme:
\begin{equation}
\langle \Sigma_\tau \rangle_{\textnormal{MH}}\in \mathbb{R},\textnormal{ whereas }
\langle \Sigma_\tau \rangle_{\textnormal{TPM}}\in \mathbb{R}^+\cup\{0\}.
\end{equation}
\end{corollary}

Let us mention that Corollary \ref{violation} is independent of the fact that the MH distribution may present negativities. What is more, the ordering of the variances of work obtained in both schemes (see Table \ref{fig:table}) cannot explain this result either (more details on these facts can be found in Appendix \ref{app_negativities}).

\section{Conclusions}
\label{conclusions}
Throughout this work we have studied the TPM and the MH distributions of work from a systematic comparative approach, being able to assess the difference between both of them in terms of quantum coherence. In particular, we have shown that the difference between the first and second moments of work obtained in both schemes is upper-bounded by the initial coherence, as quantified by the $l_{1}$-coherence measure. Regarding the variances of work, we have proven that it is not possible to establish which one is larger in general, since their difference is fundamentally sensitive to the specific configuration of the experiment. Moreover, when restricting to a specific qubit setting, the difference between variances can again be cast via the $l_1$-coherence of the initial state. This holds as well for the average entropy production, which in addition can take negative values, contrary to what is prescribed in the TPM framework. 

Our work sheds light on the formal connection between the theory of quantum coherence and recent attempts at going beyond the limitation of the TPM to unveil the statistics of energy fluctuations resulting from quantum processes. Such connection, which is becoming increasingly apparent in light of recent work~\cite{Santos2019,FrancicaPRE2019,Francica2020,Francica2019bis,Francica2020bis}, is likely to embody the {\it leit motif} of future endeavours aimed at pinpointing the potential advantages of quantum (thermo-)devices. 

\section*{Acknowledgements}
MP is grateful to Alessio Belenchia, Stefano Gherardini, Gabriel Landi, and Andrea Trombettoni for fruitful discussions and thanks Mark Mitchison and the organisers of the {\it ``Quarantine Thermo"} seminars series for giving him the opportunity to present some of the results reported here. MGD acknowledges support from Spanish MINECO reference FIS2016-80681-P (with the support of AEI/FEDER,EU) and the Generalitat de Catalunya, project CIRIT 2017-SGR-1127.  GG acknowledges support from the European Research Council Starting Grant ODYSSEY (grant nr.~758403). MP acknowledges support from the H2020-FETOPEN-2018-2020 project TEQ (grant nr.~766900),
the DfE-SFI Investigator Programme (grant 15/IA/2864),
COST Action CA15220, the Royal Society Wolfson Research Fellowship
(RSWF\textbackslash R3\textbackslash183013),
the Royal Society International Exchanges Programme
(IEC\textbackslash R2\textbackslash192220),
the Leverhulme Trust Research Project Grant (grant nr.~RGP-2018-266), and the UK EPSRC. 

\bibliographystyle{apsrev4-1}
\bibliography{thermoMP}

%%%%%%%%%%%%%%%%%%%%%%%%%%%%%%%%%%%%%%%%%%%%%%%%%%%%%%%%%%%%%%%%%%%%%%%%%%%%%%%%%%%%%%%

%%%%%%%%%%%%%%%%%%%%%%%%%%%%%%%%%

\clearpage
\onecolumngrid
\appendix

\setcounter{figure}{0}
\renewcommand{\thefigure}{A\arabic{figure}}

\section{Proof of Theorem \ref{bound_qudit}}
\label{appA}
Here we provide details on the steps to go through in order to prove the statement made in Theorem {\bf 1}. We provide such details by addressing Eqs.~\eqref{qudit_maxU} and \eqref{max_qubit} independently. 
\begin{itemize}
\item  \textbf{Eq.~(\ref{qudit_maxU})}: 
The parameterization of qudit states and unitaries makes finding an exact expression for the absolute difference between average works a difficult task to tackle. However, one can still find an upper bound to such difference as follows
\begin{equation}
\begin{aligned}
|\langle w_\tau \rangle_{\text{MH}}-\langle w_\tau \rangle_{\text{TPM}}|&=  \left | \sum_{i\neq j} \rho_{ij}  \bra{j}  U_\tau^\dagger \sum_k h_k \ketbra{k}{k} U_\tau \ket{i} \right | \\  
&\leq  \sum_{i\neq j}  | \rho_{ij} | \sum_{k}| h_k|| \bra{j}  U_\tau^\dagger \ketbra{k}{k} U_\tau \ket{i} |\\%  \label{a} \\
&\leq  \frac{1}{2}  \sum_{i\neq j}| \rho_{ij}|  \sum_k |h_k| %\label{b}\\
=\frac{1}{2} \tr|H| C_{l_1}(\rho_0),
\end{aligned}
\end{equation} 
where we have used the triangle inequality and the fact that %Note that the second line of (\ref{a}) is given by the triangle inequality and (\ref{b}) by the fact that the 
the coherence of the pure state $U_\tau^\dagger \ketbra{k}{k} U_\tau$ can never be larger than ${1}/{2}$ to achieve the final upper bound. 

 \item \textbf{Eq.~(\ref{max_qubit})}: When restricting our attention to qubits, we can parameterize unitary operations as $U_\tau=e^{i\frac{\varphi}{2}}\begin{pmatrix}
  e^{i \varphi_1} \cos \tau & e^{i \varphi _2} \sin \tau \\
  -e^{-i \varphi_2} \sin \tau & e^{-i \varphi_1} \cos \tau
  \end{pmatrix}$, which generalizes Eq.~\eqref{Umat}. This gives us %   can now give the details of the derivation of Eq.~(\ref{qubit_dif}). We have
\begin{equation}
\begin{aligned}
|\langle w_\tau \rangle_{\text{MH}}-\langle w_\tau \rangle_{\text{TPM}}|&=  \left | \sum_{i\neq j} \rho_{ij}  \bra{j}  U_\tau^\dagger \sum_k h_k \ketbra{k}{k} U_\tau \ket{i} \right | =\left| \sum_{i\neq j}\rho_{ij}\sum_{k=0,1}h_k \gamma_{ji}^{(k)}  \right| = \left| \sum_{i\neq j}\rho_{ij}\left(h_0 \gamma_{ji}^{(0)}-h_1\gamma_{ji}^{(0)}\right)  \right| \\
&= |h_0-h_1|\left| \rho_{01}\gamma_{10}^{(0)}+\rho_{10}\gamma_{01}^{(0)}  \right| = \frac{|h_0-h_1|}{2}\left|\sin(2\tau)\sqrt{a_x^2+a_y^2}\cos\left(\arctan \frac{a_y}{a_x}+\varphi_2-\varphi_1\right)\right|\\
&= \frac{|h_0-h_1|}{2}\left|\sin(2\tau)||\cos(\chi+\varphi_2-\varphi_1)\right|C_{l_1}(\rho_0)= \frac{\tr |H|}{2}|\sin(2\tau)|C_{l_1}(\rho_0)|\cos(\chi+\varphi_2-\varphi_1)|,
\end{aligned}
\end{equation}
where $\gamma^{(k)}:=U_\tau^\dagger \ketbra{k}{k} U_\tau$ and we have used that, for qubit unitaries, $\gamma^{(1)}_{ji}=-\gamma^{(0)}_{ji}$.
%\item \textbf{Eq. (\ref{max_qubit})}:
For a fixed Hamiltonian $H$, such difference is maximized by choosing $\varphi_2-\varphi_1=-\chi$ and $\tau=\pi/4$. 
\end{itemize}

\section{Proof of Theorem \ref{qudit_second}}
\label{appB}
We now pass to the proof of the statement in Theorem~\ref{qudit_second}, for which we need to go through the following steps. 
\begin{itemize}
\item \textbf{Eq. (\ref{bound_qudit_second})}: By using the triangle inequality, the definition of $\gamma^{(k)}$ given above and the fact that $\gamma^{(k)}_{ji}\le1/2$, we can also provide the following upper bound %to $|\langle  w^2 \rangle_{\text{MH}}-\langle  w^2 \rangle_{\text{TPM}}|$ as follows
\begin{equation}
\begin{aligned}
|\langle  w^2_\tau \rangle_{\text{MH}}-\langle  w^2_\tau \rangle_{\text{TPM}}|&\leq  \sum_{i\neq j} |\rho_{ij}| \sum_k |h_k^2| |\gamma^{(k)}_{ji}|+\sum_{i\neq j} |\rho_{ij}| \sum_l |h_l| |h_i| |\gamma^{(l)}_{ji}| +\sum_{i\neq j} |\rho_{ij}| \sum_m |h_j| |h_m| |\gamma^{(m)}_{ji}| \\
&\leq \frac{1}{2}C_{l_1}(\rho_0) \tr H^2 +\frac{1}{2}C_{l_1}(\rho_0)\max_k |h_k|\tr|H|+\frac{1}{2}C_{l_1}(\rho_0)\max_k |h_k|\tr|H|.
\end{aligned}
\end{equation}
%where we have used the triangle inequality, the usual definition $\sigma^{(k)}:=U(\tau)^\dagger \ketbra{k}{k} U(\tau)$ and the fact that $\sigma^{(k)}_{ji}$ is always smaller than $\frac{1}{2}$.\\

\item \textbf{Eq. (\ref{second_qubit})}: When focusing on qubits we have
\begin{equation}
\begin{aligned}
|\langle  w^2_\tau \rangle_{\text{MH}}-\langle  w^2_\tau \rangle_{\text{TPM}}|&=  \left | \sum_{i\neq j} \rho_{ij} \bra{j} U_\tau^\dagger H^2 U_\tau - U_\tau^\dagger HU_\tau H - H U_\tau^\dagger HU_\tau \ket{i} \right |\\
&=\left| \sum_{i\neq j}\rho_{ij} \left(\sum_k h_k^2 \gamma^{(k)}_{ji}-\sum_l h_l h_i \gamma^{(l)}_{ji} -\sum_m h_j h_m \gamma^{(m)}_{ji} \right) \right| \\
&= \left| \sum_{i\neq j}\rho_{ij}\left[\gamma^{(0)}_{ji}(h_0^2-h_0h_i-h_jh_0)+\gamma^{(1)}_{ji}(h_1^2-h_1h_i-h_j h_1)\right]\right| \\
&= \left| \sum_{i\neq j}\rho_{ij} \gamma^{(0)}_{ji}\left(h_0^2-h_0h_i -h_jh_0-h_1^2+h_1h_i+h_jh_1\right)   \right|\\
&= \left| (\rho_{01} \gamma_{10}^{(0)}+\rho_{10} \gamma_{01}^{(0)})(h_0^2-h_0^2-h_1h_0-h_1^2+h_1h_0+h_1^2)\right|=0.
\end{aligned}
\end{equation}
%where $\sigma^{(k)}$ is the pure state $\sigma^{(k)}:=U(\tau)^\dagger \ketbra{k}{k} U(\tau)$  and where we have used that, for qubit unitaries, $\sigma^{(1)}_{ji}=-\sigma^{(0)}_{ji}$.
\end{itemize}

\section{Derivation of Table \ref{fig:table}}\label{app_variances}
We now give an assessment of the relations reported in Table~\ref{fig:table}. The first thing to notice is that $( \Delta w_\tau)^2_{\textnormal{MH}}- ( \Delta w_\tau)^2_{\textnormal{TPM}}$ has roots at $a_x=0$ and $a_x=-2a_z\tan \tau$, which means that there are two points at which the variances coincide. The first one comes from the equivalence between both schemes when we consider vanishing initial coherence. Moreover, for $a_x=\pm 1$ we have $( \Delta w_\tau)^2_{\textnormal{MH}}- ( \Delta w_\tau)^2_{\textnormal{TPM}}=-(h_0-h_1)^2\cos(\tau)^2\sin(\tau)^2<0$. Let us now consider $a_z> 0$ and $\tan\tau> 0$. First, due to Bolzano's theorem, the difference between variances for $a_x\in[-1,-2a_z\tan\tau]$ has to be negative: as it is already negative at $a_x=-1$, having a positive difference within such interval would mean that there should be another root inside it, which is not the case. Second, for the same reason, the difference between variances should have a fixed sign for $a_x\in[-2a_z\tan\tau,0]$. In particular, such difference must be positive as 
\begin{equation}
\left.\dfrac{\partial[ ( \Delta w_\tau)^2_{\textnormal{MH}}- ( \Delta w_\tau)^2_{\textnormal{TPM}}]}{\partial a_x}\right\vert_{a_x=-2a_z\tan\tau}=2(h_0-h_1)^2a_z\sin(\tau)^4\left(\dfrac{1}{\tan\tau}+4\tan\tau\right)>0.
\end{equation}
Finally, we have that the difference between variances is negative for $a_x\in [0,1]$, again due to Bolzano's theorem. The same arguments can be applied to the rest of the cases, i.e. $\text{sgn}(a_z)=\text{sgn}(\tan\tau)$ and $\text{sgn}(a_z)\neq \text{sgn}(\tan\tau)$. 

\section{Proof of Theorem \ref{mh_lr}}
Let us now move to the proof of Theorem~\ref{mh_lr}.
\begin{itemize}
\item \textbf{Eq. (\ref{entropylr})}:
The Jarzynski equality~\cite{jarzynski1997} $\langle e^{-\beta (w_\tau-\Delta F_\tau)}\rangle_{\text{TPM}}=1$ is only fulfilled when the initial state is at equilibrium.
For an arbitrary initial state $\rho$, the following fluctuation theorem applies~\cite{allahverdyan2014}
\begin{equation}
\langle e^{-\beta (w_\tau-\Delta F_\tau)}\rangle_{\text{MH}}= \re\left(\tr[\gamma_\tau {\cal G}_0^{-1}\rho_0]\right) \equiv \xi_\tau,
\end{equation}
where ${\cal G}_\lambda=\dfrac{e^{-\beta H_\lambda}}{\tr [e^{-\beta H_\lambda}]}$ is a Gibbs state and $\gamma_\tau=U_\tau^\dagger {\cal G}_\tau U_\tau$. From here we get that the free energy difference in the MH scheme is given by
\begin{equation}
\Delta F_\tau=-(\ln \langle e^{-\beta w_\tau}\rangle_{\text{MH}}-\ln \xi_\tau )/\beta.
\end{equation}
We use this result in the definition of entropy production $\Sigma=\beta(w-\Delta F)$ and use a cumulant expansion of $\langle e^{-\beta w_\tau}\rangle_{\text{MH}}$  to find ~\cite{Batalhao2015}
\begin{equation}
\langle \Sigma_\tau \rangle_{\text{MH}}= \sum_{n\geq 2} \frac{(-1)^n}{n!} \kappa_\tau^{(n)}(\beta) \beta^n -\ln \xi_\tau,
\end{equation}
where $\kappa_\tau^{(n)}$ are the cumulants of the MH work distribution. Note that we do not take the average of $\ln \xi_\tau$, as it does not contain any stochastic variable $w_\tau$. 

In the linear response regime, the first term yields $\frac{\beta^2}{2} (\Delta w_\tau)^2_{\textnormal{MH}}$ \cite{Batalhao2015}, where $(\Delta w_\tau)^2_{\textnormal{MH}}=\kappa_\tau^{(2)}(\beta)$ is the variance of the MH distribution of work. Expanding the second term gives \MP{}
\begin{equation}
\begin{aligned}
\ln\xi_\tau %&\approx \ln \dfrac{Z(\lambda_0)}{Z(\lambda_\tau)}+\ln (1-\beta \langle w \rangle_{\textnormal{MH}}+\dfrac{\beta^2}{2}(\langle w^2 \rangle_{\textnormal{MH}}+\re \tr \rho [H(\lambda_0),U(\tau)^\dagger H(\lambda_\tau)U(\tau)]))\\
&\approx\frac{\beta^2}{4}\tr[ H_0^2- H_\tau^2]-\beta \langle w_\tau \rangle_{\text{MH}}+\frac{\beta^2}{2}\left\{(\Delta w_\tau)^2_{\text{MH}}+\re\left( \tr[ \rho_0 [H_0,U_\tau^\dagger H_\tau U_\tau]]\right)\right\}.
\end{aligned}
\end{equation}
Thus, the average entropy production in the linear response regime amounts to
\begin{equation}
\langle \Sigma_\tau \rangle_{\textnormal{MH}}^{\textnormal{LR}}=\beta \langle w_\tau\rangle_{\textnormal{MH}}-\dfrac{\beta^2}{2}\re\left( \tr [\rho_0 [H_0,U_\tau^\dagger H_\tau U_\tau]]\right) -\dfrac{\beta^2}{4}\tr[ H_0^2-H_\tau^2].
\end{equation} 

\item \textbf{Eq. (\ref{entropylrqubit})}:
Let us now have a close look at qubits. For convenience, we consider a qubit prepared in the state $\rho=\frac{1}{2}\begin{pmatrix}
1-a_z & a_x-i a_y \\
a_x+i a_y & 1+a_z 
\end{pmatrix}$,
with $\vec{a} \in \mathbb{R}^3$,  $\dfrac{1-a_z}{2}=\dfrac{e^{-\beta}}{\tr[e^{-\beta H_0}]}$ and $H_0=\sigma_z$. This ensures that, for small enough $\beta$, $a_z$ will also be small: 
\begin{equation}
\dfrac{1-a_z}{2}\approx \dfrac{1-\beta}{2}\rightarrow a_z\approx \beta.
\end{equation}

Let us suppose the qubit is subjected to a real unitary transformation $U_\tau$ such that $H_\tau=k\sigma_z$, for $k\in \mathbb{C}$. The average entropy production in the linear response regime is then given by Eq.~(\ref{entropylr})
\begin{equation}
\begin{aligned}
\langle \Sigma_\tau \rangle_{\textnormal{MH}}^{\textnormal{LR}}&=a_x\beta k \sin(2\tau)-2a_z\beta k\cos(\tau)^2+\frac{\beta^2 k^2}{2}+a_z\beta k-\frac{\beta^2}{2}+a_z \beta\\
&\approx \beta k\sin(2\tau)\cos(\chi)C_{l_1}(\rho_0)-2\beta^2 k\cos(\tau)^2+\frac{\beta^2 k^2}{2}+\beta^2 k-\frac{\beta^2}{2}+\beta^2\\
&= \beta k \sin(2\tau)\cos(\chi)C_{l_1}(\rho_0)-2\beta^2 k\cos(\tau)^2+\frac{\beta^2 k^2}{2}+\beta^2 k+\frac{\beta^2}{2},
\end{aligned}
\end{equation}
where we have used that, for small $\beta$, $a_z\approx \beta$. As shown in Ref.~\cite{Batalhao2015}, the TPM average entropy production in the linear response regime, where the initial state is set to be in equilibrium, is given by $\langle \Sigma_\tau \rangle_{\textnormal{TPM}}^{\textnormal{LR}}=\frac{\beta^2}{2} ( \Delta w_\tau)^2_{\textnormal{TPM}}$. Let us compute it for our qubit evolution
\begin{equation}
\begin{aligned}
\langle \Sigma_\tau \rangle_{\textnormal{TPM}}^\text{LR}&= 2 \beta^2 k (1-\cos(\tau)^2)-2 a_z^2 \beta^2 k^2 \sin(\tau)^4+2a_z^2 \beta^2 k^2\sin(\tau)^2\\
&-2 a_z^2 \beta^2 k\sin(\tau)^2-\frac{a_z^2 \beta^2 k^2}{2}+\frac{\beta^2 k^2}{2}+a_z^2\beta^2 k-\beta^2 k-\frac{a_z^2 \beta^2}{2}+\frac{\beta^2}{2}\\
&\approx-2\beta^2 k \cos(\tau)^2+\frac{\beta^2 k^2}{2}+\beta^2 k+\frac{\beta^2}{2},
\end{aligned}
\end{equation}
where we have neglected the terms in $a_z^2\beta^2\approx \beta^4$. Therefore 
\begin{equation}
\langle \Sigma_\tau \rangle_{\textnormal{MH}}^{\textnormal{LR}}=\beta k \sin(2\tau)\cos(\chi)C_{l_1}(\rho_0)+ \langle \Sigma_\tau \rangle_{\textnormal{TPM}}^{\textnormal{LR}}.
\end{equation}

\end{itemize}

\begin{figure}[t!]

\subfloat[\label{fig:entropygeneralbeta02}]{%
	\includegraphics[width=.52\linewidth]{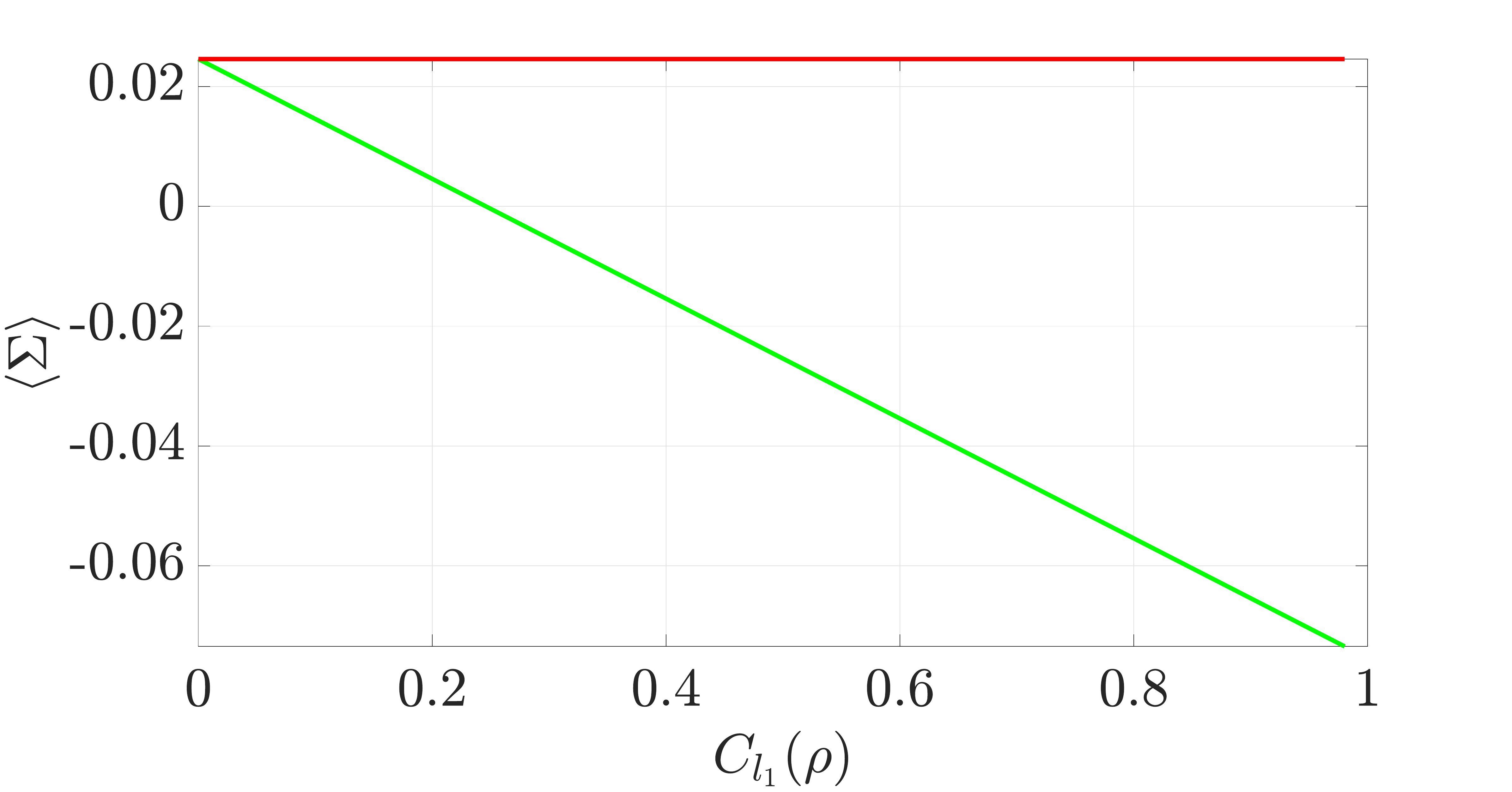}%
}
\subfloat[\label{fig:entropydecomposedbeta02}]{%
	\includegraphics[width=.52\linewidth]{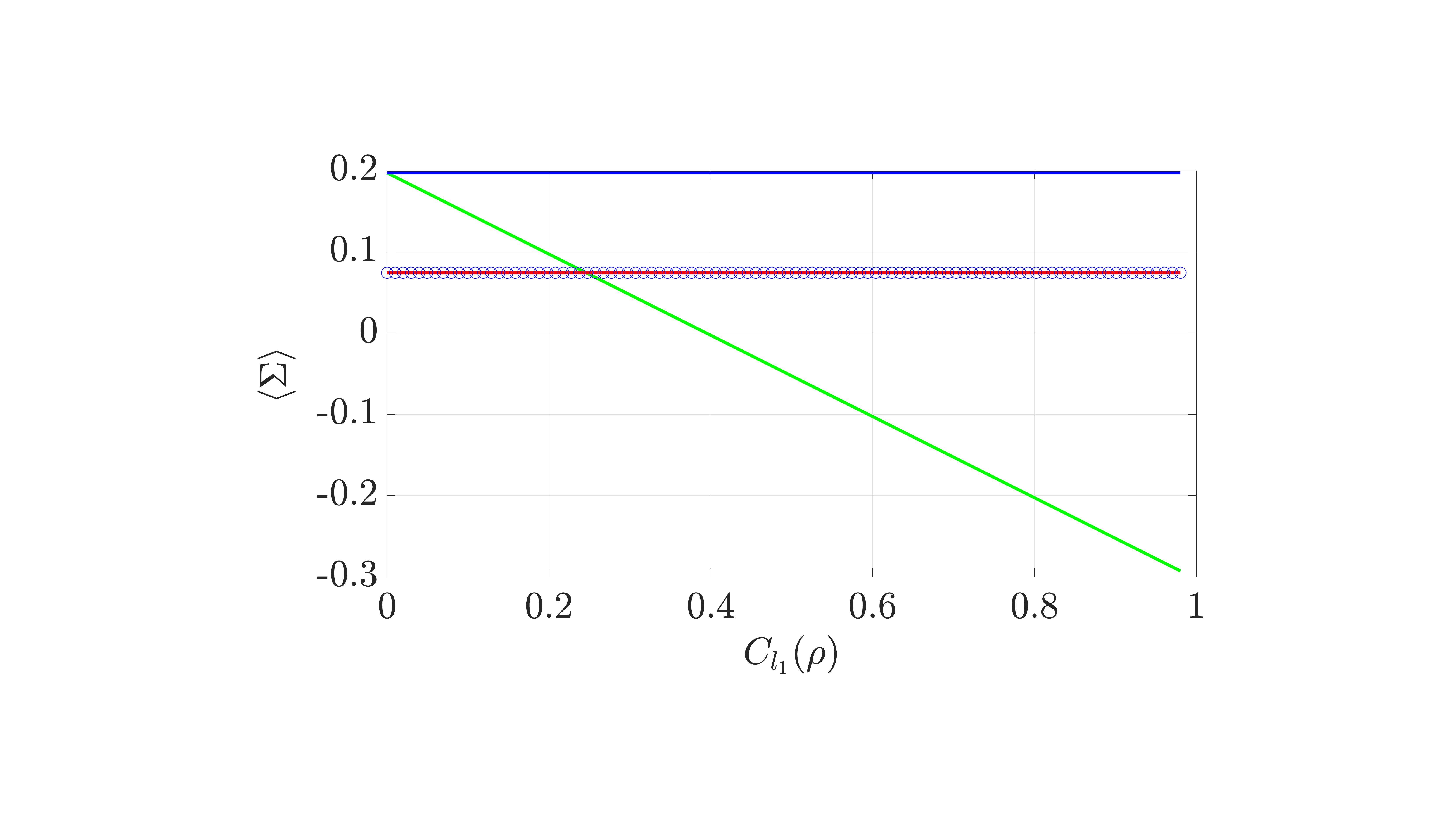}%
}
\\
\subfloat[\label{fig:entropynegativity}]{%
	\includegraphics[width=.52\linewidth]{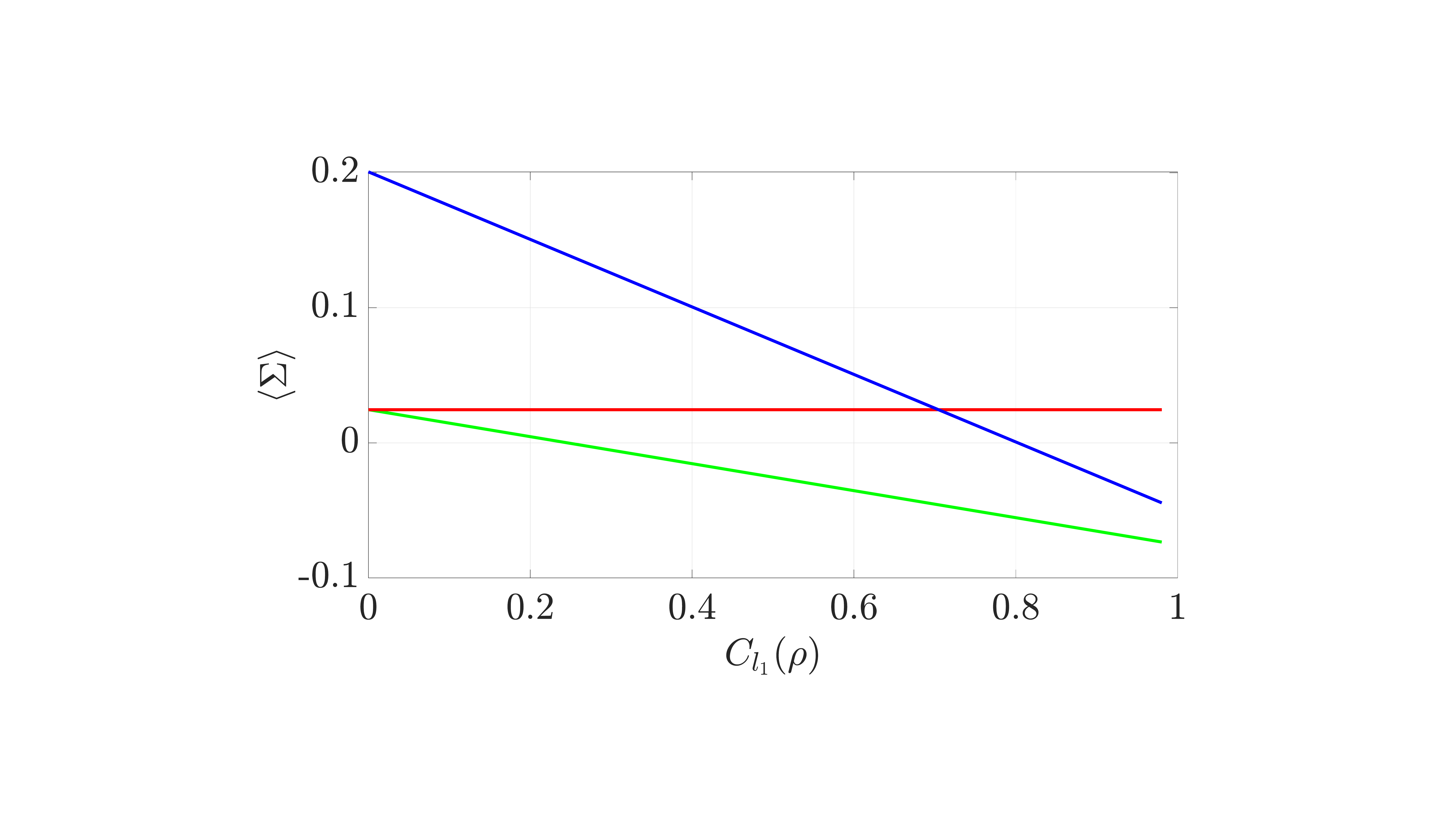}%
}
\subfloat[\label{fig:entropyinsidelog}]{%
	\includegraphics[width=.52\linewidth]{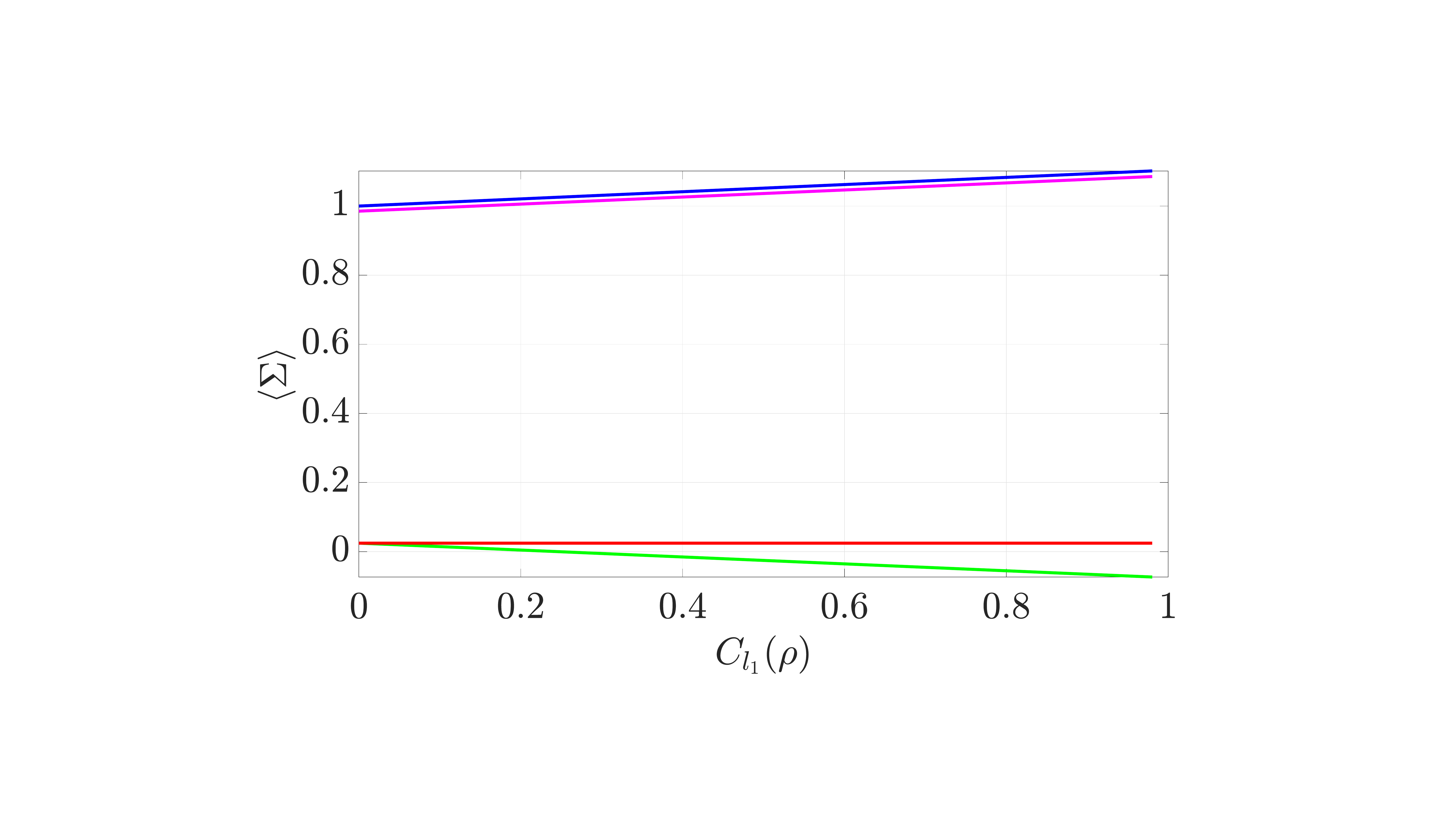}%
}
\caption{Qubit in the initial state $\rho_0=\begin{pmatrix}
	1-\alpha^2 & \omega \alpha \sqrt{1-\alpha^2}  \\
	\omega \alpha \sqrt{1-\alpha^2}  & \alpha^2 
	\end{pmatrix}$, 
	where $\alpha^2=\dfrac{e^{\beta }}{\tr e^{-\beta H(\lambda_0)}}$, $0\leq \omega\leq 1$ and $H(\lambda_0)=\sigma_z$, undergoing an evolution given by a real unitary  
	$U_\tau=\begin{pmatrix}
	\cos(\tau) & \sin(\tau) \\
	-\sin(\tau) & \cos(\tau) 
	\end{pmatrix}$
	and $H(\lambda_\tau)=\frac{1}{2}\sigma_z$.
	\textbf{a)} $\langle \Sigma_\tau \rangle$ versus initial coherence, $\beta=0.2$, $\tau=\frac{3\pi}{4}$. MH scheme (green) and TPM scheme (red).
	\textbf{b)} $\langle w_\tau \rangle_{\text{MH}}$ (green), $\langle w_\tau \rangle_{\text{TPM}}$ (blue line), $\Delta F_{\tau,\text{MH}}$ (red) and $\Delta F_{\tau,\text{TPM}}$ (blue circles) (naturally, they both agree),  versus initial coherence, $\beta=0.2$, $\tau=\frac{3\pi}{4}$. 
	\textbf{c)} $\langle \Sigma_\tau \rangle$ versus initial coherence, $\beta=0.2$, $\tau=\frac{3\pi}{4}$. MH scheme (green) and TPM scheme (red). Negativity of the MH distribution, computed as $\min_{mn} \re \tr ( U_\tau^\dagger \ketbra{n}{n}U_\tau\ketbra{m}{m}\rho_0) $ (blue).
	\textbf{d)} $\langle \Sigma_\tau \rangle$ versus initial coherence, $\beta=0.2$, $\tau=\frac{3\pi}{4}$. MH scheme (green) and TPM scheme (red). $\langle e^{-\beta w_\tau} \rangle_{\text{MH}}$ (magenta) and $\xi_\tau$ (blue).
}
\label{fig:entropy_general}
\end{figure}

\section{Proof of Corollary \ref{violation}}
Consider the qubit case for which Eq. (\ref{entropylrqubit}) holds. For $k=\frac{1}{2}$, $\tau=\frac{3\pi}{4}$ and $\chi=0$, we have that
\begin{equation}
\langle \Sigma_\tau \rangle_{\textnormal{MH}}^{\textnormal{LR}}=-\frac{\beta}{2}C_{l_1}(\rho_0) +\frac{5\beta^2}{8},
\end{equation} 
which becomes negative for $C_{l_1}(\rho_0)>\frac{5\beta}{4}$. For example, for $\beta=0.2$, the average entropy production is negative when $C_{l_1}(\rho_0)>0.25$, as it is shown in Fig. \ref{fig:entropygeneralbeta02}.   

\section{Why can $\langle\Sigma_\tau \rangle_{\text{MH}}$ be negative?}\label{app_negativities}
Let us look closely into the case where $\beta \rightarrow 0$. In Fig. \ref{fig:entropygeneralbeta02} we can see how the average entropy production changes with the initial coherence. As expected, in the TPM scheme the entropy remains constant, while in the MH scheme it decreases, being even able to take negative values. The same can be noticed in Fig. \ref{fig:entropyvstimebeta02}. Indeed, for small enough $\beta$, the average work calculated in the MH framework can be smaller than the free energy, and even negative (cf. Fig. \ref{fig:entropydecomposedbeta02}). This immediately suggests why the average entropy production can be negative in the MH scheme: while $\Delta F_\tau$ is not sensitive to coherence and thus remains constant, $\langle w_\tau \rangle_{\text{MH}}$ keeps on decreasing as the initial coherences increase. 

What is more, this fact does not seem to be due to the MH presenting negativities: in Fig. \ref{fig:entropynegativity} we see that the violation may persist under a non-negative MH distribution. Besides, non-negativities still allow for well-defined logarithms $\ln \langle e^{-\beta w_\tau} \rangle_{\text{MH}}$ and $\ln \xi_\tau$, since $\langle e^{-\beta w_\tau} \rangle_{\text{MH}}$ and $\xi_\tau$ are positive, respectively  (see Fig. \ref{fig:entropyinsidelog}). Moreover, the ordering between the variances of work obtained in the MH and the TPM schemes does not seem to provide an explanation on why the average entropy production can be negative in the MH scheme: according to Table \ref{fig:table}, the higher uncertainty of the MH scheme compared to that of the TPM scheme ($(\Delta w_\tau)^2_{\text{MH}}\geq (\Delta w_\tau)^2_{\text{TPM}}$) would occur within the interval $0\leq C_{l_1}(\rho_0)\leq 0.4$. However, the average entropy production in that interval can take any sign. Furthermore, $(\Delta w_\tau)^2_{\text{MH}}\leq (\Delta w_\tau)^2_{\text{TPM}}$ holds for $0.4\leq C_{l_1}(\rho_0)\leq 1$, where the average entropy production is always negative (see Fig. \ref{fig:entropygeneralbeta02}).

\end{document}